\documentclass[]{emulateapj}

\usepackage{graphicx}
\usepackage{color,soul}
\usepackage{natbib}
\usepackage{amsmath}

\begin{document}

\title{Constraining the Milky Way's Hot Gas Halo with \ion{O}{7} and \ion{O}{8} Emission Lines}
\author{Matthew J. Miller and Joel N. Bregman}
\affil{Department of Astronomy, University of Michigan, Ann Arbor, MI 48104, USA}
\email{mjmil@umich.edu, jbregman@umich.edu}

\begin{abstract}

The Milky Way hosts a hot ($\approx 2 \times 10^6$ K), diffuse, gaseous halo based on detections of z = 0 \ion{O}{7} and \ion{O}{8} absorption lines in quasar spectra and emission lines in blank-sky spectra.  Here we improve constraints on the structure of the hot gas halo by fitting a radial model to a much larger sample of \ion{O}{7} and \ion{O}{8} emission line measurements from \textit{XMM-Newton}/EPIC-MOS spectra compared to previous studies ($\approx$650 sightlines). We assume a modified $\beta$-model for the halo density distribution and a constant-density Local Bubble from which we calculate emission to compare with the observations.  We find an acceptable fit to the \ion{O}{8} emission line observations with $\chi^{2}_{red}$ (dof) = 1.08 (644) for best-fit parameters of $n_o r_c^{3\beta} = 1.35 \pm 0.24$ cm$^{-3}$ kpc$^{3\beta}$ and $\beta = 0.50 \pm 0.03$ for the hot gas halo and negligible Local Bubble contribution.  The \ion{O}{7} observations yield an unacceptable $\chi^{2}_{red}$ (dof) = 4.69 (645) for similar best-fit parameters, which is likely due to temperature or density variations in the Local Bubble.  The \ion{O}{8} fitting results imply hot gas masses of $M$($\textless 50$ kpc) = $3.8_{-0.3}^{+0.3} \times 10^{9} M_{\odot}$ and $M$($\textless 250$ kpc) = $4.3_{-0.8}^{+0.9} \times 10^{10} M_{\odot}$, accounting for $\lesssim$50\% of the Milky Way's missing baryons.  We also explore our results in the context of optical depth effects in the halo gas, the halo gas cooling properties, temperature and entropy gradients in the halo gas, and the gas metallicity distribution.  The combination of absorption and emission line analyses implies a sub-solar gas metallicity that decreases with radius, but that also must be $\geq 0.3 Z_{\odot}$ to be consistent with the pulsar dispersion measure toward the Large Magellanic Cloud.

\end{abstract}

\section{Introduction}
\label{section.introduction}

The detection of X-ray-emitting and -absorbing gas at zero redshift implies the existence of a diffuse volume-filling plasma associated with the Milky Way's interstellar medium (ISM) and circumgalactic medium (CGM).  Detections of these absorption and emission lines along individual lines of sight imply plasma densities between $10^{-5}$ and $10^{-3}$ cm$^{-3}$ and plasma temperatures of $\sim 10^{6}$ K.  These observations constrain the detailed properties of the gas along individual sightlines, but there has not been a comprehensive analysis on the global properties of the absorption and emission lines.  Analyzing the global properties of the observations is necessary to constrain the overall structure and extend of the gas.  Constraints on the structure of the gas, specifically the density profile, are necessary to estimate the hot gas mass within the Milky Way's virial radius.  The contribution of the hot gas to the Milky Way's baryon budget may account for a significant fraction of the Milky Way's ``missing baryons''.  

The tracers of the Milky Way's hot halo gas are \ion{O}{7} and \ion{O}{8} absorption and emission lines characteristic of gas in the 10$^6$ - 10$^7$ K range \citep{paerels_kahn03}.  The emission lines are thought to be a significant contributor to the 0.5 - 2.0 keV portion of the Milky Way's diffuse soft X-ray background \citep[SXRB; ][]{snowden_etal97, mccammon_etal02, hs12} and are typically seen in otherwise empty fields of view in the sky \citep[$\sim$1000 sightlines; ][]{yoshino_etal09, hs10, hs12}.  On the other hand, the absorption lines are only seen in $\sim$30 bright active galactic nucleus (AGN) and blazar spectra \citep{nicastro_etal02, rasmussen_etal03, wang_etal05, williams_etal05, fang_etal06, bregman_ld07, yao_wang07, hagihara_etal10, gupta_etal12, yao_etal12, miller_bregman13, fang_jiang14} or X-ray binaries \citep{yao_wang05, hagihara_etal10}.  The sensitivity of current X-ray telescopes in the soft X-ray band is the limiting factor on the number of absorption line detections, but still results in the nearly ubiquitous detection of the emission lines.  

The SXRB, with some contribution from \ion{O}{7} and \ion{O}{8} emission lines, has been observed in broadband images from the \textit{ROSAT} All-Sky Survey \citep[RASS; ][]{snowden_etal97}, in blank-sky spectra from current X-ray telescopes \citep{yoshino_etal09, hs10, hs12}, and with the Diffuse X-ray Spectrometer sounding rocket \citep{mccammon_etal02}.  A combination of results from analyses comparing the RASS 1/4 keV and 3/4 keV bands \citep{snowden_etal97, kuntz_snowden00} and shadowing experiments toward nearby molecular clouds \citep{galeazzi_etal07, smith_etal07} imply that multiple plasma components comprise the 0.5 - 2.0 keV component of the SXRB.  

The ``local'' emission component of the SXRB is believed to come from a combination of the Local Bubble (LB) and solar wind charge exchange (SWCX) processes.  The LB is thought to be a supernova (SN) remnant that the Sun is currently inside \citep{snowden_etal90, snowden_etal93}, although its physical and corresponding emission properties are debated in the literature.  Arguments exist either for the LB being filled with X-ray-emitting gas at $\sim 10^{6}$ K \citep{smith_etal07} or that the emission comes more from a wall of material at the edges of the bubble \citep[100-300 pc away; e.g.,][]{welsh_shelton09}.  On the other hand, SWCX emission is a known soft X-ray emission source where neutral hydrogen and helium atoms undergo charge exchange reactions with solar wind ions in and around our solar system \citep{cravens_etal01, snowden_etal04, wargelin_etal04, carter_sembay08, koutroumpa_etal07, kuntz_snowden08, carter_etal11, ezoe_etal10, koutroumpa_etal11}.  These reactions are difficult to predict or quantify, but are known to produce time-variable line emission at energies $\lesssim$1 keV.  While the details of both emission sources are still unclear, a combination of both SWCX and LB emission models are necessary to reproduce the \textit{ROSAT} 1/4 keV band emission \citep{galeazzi_etal14, smith_etal14}.  The picture at high energies, specifically for the \ion{O}{7} and \ion{O}{8} emission lines, has uncertainties as well.  Shadowing experiments toward nearby molecular clouds show that local \ion{O}{7} emission is common, but \ion{O}{8} emission is not always detected \citep{smith_etal07, koutroumpa_etal11}.  It is clear the LB and/or SWCX \textit{can} produce \ion{O}{7} and \ion{O}{8} emission, but their global oxygen line emission properties are still unclear.  

The ``non-local'' plasma component of the SXRB is believed to come from a more extended, diffuse plasma at a slightly hotter temperature than the LB \citep{yoshino_etal09, hs13}, although the source and exact spatial extent of the plasma is unclear.  One potential source would be a plane-parallel, exponential distribution of $\sim 10^{6}$ K due to SN-driven outflows from the Milky Way's disk \citep{norman_ikeuchi89, joung_maclow06, hill_etal12}.  Indeed, this type of density distribution is suggested by \cite{hagihara_etal10} based on \ion{O}{7} and \ion{O}{8} absorption line measurements along the PKS 2155-304 sightline.  They fit their absorption line observations with an exponential disk density model with a scale height of $2.8_{-1.0}^{+6.4}$ kpc, implying a much more confined medium compared to the diffuse, volume-filled halo picture.  However, this type of distribution results in inconsistencies with additional ``non-local'' plasma observables, including the emission measure distribution across the sky of the ``non-local'' plasma \citep{henley_etal10, hs13} and the Milky Way's diffuse X-ray surface brightness \citep{fang_etal13}.  This implies that the ``non-local'' plasma contribution to the SXRB, and thus the \ion{O}{7} and \ion{O}{8} emission lines, is likely from an extended distribution of gas in the Milky Way's halo.  

Analyses of the \ion{O}{7} and \ion{O}{8} absorption lines typically assume that the lines arise from a large (scales $\gtrsim$20 kpc), diffuse, volume-filled halo consistent with shock heated gas at the Milky Way's virial temperature in quasi-static equilibrium \citep{white_frenk91}.  Detailed work on the absorption strengths between \ion{O}{7} and \ion{O}{8} (when detected) along individual sightlines suggests a plasma temperature between log($T$) = 6.1 - 6.4, but do not provide constraints on the large-scale properties of the plasma.  Recent work by \cite{bregman_ld07} and \cite{gupta_etal12} analyzed larger samples of the absorption lines using \textit{XMM-Newton} Reflection Grating Spectrometer and \textit{Chandra} High and Low Energy Transmission Grating data.  \cite{bregman_ld07} compared their column densities to a single line of sight emission measurement \citep{mccammon_etal02} while \cite{gupta_etal12} compared their column densities to an average emission measure from $\sim$20 lines of sight \citep{yoshino_etal09, henley_etal10} to estimate a characteristic density and path length of the absorbing material.  There are several discrepancies between these works, but both find characteristic densities between $10^{-4}$ and $10^{-3}$ cm$^{-3}$ and path lengths $\textgreater$20 kpc.  

\cite{miller_bregman13} provided an improvement on these works by modeling the local \ion{O}{7} absorption lines in 29 AGN and X-ray binary spectra with a more physical hot halo density model (as opposed to a constant-density sphere).  They found the absorption lines could be modeled with a modified spherical $\beta$-model (effectively a power law) with $\beta$ ranging between 0.56 and 0.71, depending on the effects of absorption line saturation ($\beta$ = 0.5 corresponds to $n \propto r^{-3/2}$).  These density model constraints resulted in hot gas mass estimates of $3.3 - 9.8 \times 10^{10} M_{\odot}$ within 200 kpc.  

Until now, there has been no comprehensive comparison between the \ion{O}{7} and \ion{O}{8} emission and absorption line observations thought to be due to a Galactic hot gas halo.  This is partially due to the difficulty in disentangling the LB and hot halo contributions to the emission lines across the entire sky.  The LB in particular is believed to be a stronger contribution to the emission lines than the absorption lines, thus further complicating any analysis.  Additionally, the differences in sample sizes between the two observables have prevented a large-scale analysis between the two.  The number of \ion{O}{7} absorption line measurements with current X-ray telescopes ($\sim$30) is at its maximum due to the sensitivity of current detectors, while many published \ion{O}{7} and \ion{O}{8} emission line measurements are focused on individual sightlines or are limited to a certain region of the sky.  This is important since the absorption and emission properties of the hot gas halo vary across the sky, meaning comparisons between absorption lines in one area of the sky with emission lines in a different area may yield incorrect results about the hot halo plasma.  

\cite{hs12} have offered a resolution to the latter issue by presenting an all-sky catalog of \ion{O}{7} and \ion{O}{8} emission line measurements of the SXRB using \textit{XMM-Newton}/EPIC-MOS data.  Their sample is an incredibly useful tool for probing the Milky Way's hot gas halo since there are many targets (1868 in their whole sample) and their sample covers the entire sky.  The combination of these two effects should provide improved constraints on the Milky Way's hot gas halo compared to the absorption line data sets, even with the known LB emission.  

In this work, we unify the procedure outlined in \cite{miller_bregman13} with this \ion{O}{7} and \ion{O}{8} emission line data set to constrain the density properties of the Milky Way's hot gas halo.  We develop a parametric model for the Milky Way's hot gas halo density profile and LB, calculate model line intensities along each line of sight and for a given parameter set, and find the model parameters that are most consistent with the data.  In this way, we constrain the density properties of the hot gas halo and place more precise estimates on important quantities such as its mass and metallicity.  

There are several advantages to this work on the \ion{O}{7} and \ion{O}{8} emission lines.  The most critical benefit of this work is that we are analyzing the emission lines in the same way \cite{miller_bregman13} analyzed \ion{O}{7} absorption lines.  This is crucial since we expect the Milky Way's hot halo to contribute to both sets of observables, so any similarities or differences between the two results can tell us about the physical properties of the gas.  The other main benefit is that the quality of our constraints using this emission line sample should be much improved compared to the absorption line constraints.  This is largely due to the increase in sample size by a factor of $\approx$20.  

The outline for the rest of the paper is as follows.  In Section ~\ref{section.data_reduction}, we describe the observation selection, data reduction, and line measurement procedure of the \ion{O}{7} and \ion{O}{8} emission lines from \cite{hs12}.  In Section ~\ref{section.model_fitting}, we describe our model fitting procedure.  This includes a discussion of our parametric model and emission line calculation.  In Section ~\ref{section.results}, we discuss our model fitting results and constraints on our density model.  In Section ~\ref{section.discussion}, we discuss the implications of our results and compare them with other studies on the Milky Way's hot gas halo.  Finally in Section ~\ref{section.summary}, we summarize our results.

\section{Data Reduction}
\label{section.data_reduction}

Our emission line sample is a subset of the comprehensive \ion{O}{7} and \ion{O}{8} emission line sample from \cite{hs12}.  The authors compiled their sample from all public \textit{XMM-Newton} observations prior to 2010 August 4 that contained any EPIC-MOS exposure time (5698 observations).  The goal of creating this sample was to analyze the various sources of the Milky Way's SXRB, including SWCX, the LB, and the Milky Way's hot gas halo.  Our data set is a subset of the \cite{hs12} full sample that maximizes our sensitivity to the Milky Way's hot halo emission.  

This work focuses on the analysis and modeling of our subset of the emission lines presented in \cite{hs12}, not the line measurements or sample compilation itself.  Here we provide an overview of their procedure and refer the reader to the full sample references for a more detailed description of the sample \citep{hs10, hs12}.  We describe their data selection and reduction procedure in Section ~\ref{subsection.data_filtering},  their emission line measurement procedure in Section ~\ref{subsection.line_measurements}, and our additional screening procedure to create our sample in Section ~\ref{subsection.our_screening}.  

\subsection{Data Filtering}
\label{subsection.data_filtering}

The 5698 archival \textit{XMM-Newton} observations were processed with the standard \textit{XMM-Newton} Science Analysis System \footnote{\url{http://xmm.esac.esa.int/sas/}} version 11.0.1, including the \textit{XMM-Newton} Extended Source Analysis Software \footnote{\url{http://heasarc.gsfc.nasa.gov/docs/xmm/xmmhp\_xmmesas.html}} \citep[\textit{XMM}-ESAS;][]{kuntz_snowden08, snowden_kuntz11}.  The \textit{XMM}-ESAS script \texttt{mos-filter} was used to remove any observing time affected by soft-proton flaring.  This contamination is identified by an excess or deficit in the 2.5-12 keV count rate.  After this filtering, the authors kept observations with $\geqslant$5 ks of good observing time and with at least one good exposure with the MOS1 and MOS2 cameras each.  This resulted in a sample of 2611 observations out of the original 5698.  

Bright X-ray sources were removed from the observations using both visual inspection and automated source removal procedures.  These screening methods are necessary to isolate SXRB emission in the extracted spectra.  The authors used data from the Second \textit{XMM-Newton} Serendipitous Source Catalog \footnote{\url{http://xmmssc-www.star.le.ac.uk/Catalogue/2XMMi-DR3/}} \citep[2XMM;][]{watson_etal09} for automated X-ray point source removal.  For each observation, any 2XMM source with 0.5 - 2.0 keV flux $F_X^{0.5-2.0} \geqslant 5 \times 10^{-14}$ erg cm$^{-2}$ s$^{-1}$ inside the field of view was removed using a 50$\arcsec$ circular region.  These regions typically enclose $\approx$90\% of the sources' fluxes.  The authors also employed a visual inspection of the observations to remove any bright or extended sources that were not included in the automated point source removal.  The exclusion regions for these sources typically ranged between 1$\arcmin$ and 4$\arcmin$ \citep{hs10}.

SWCX reactions are a known source of \ion{O}{7} and \ion{O}{8} line emission that must be accounted for in the emission line measurements.  These reactions occur between solar wind ions and neutral hydrogen atoms in the Earth's atmosphere (geocoronal SWCX) or neutral hydrogen and helium atoms in the heliosphere (heliospheric SWCX).  Geocoronal SWCX emission is strongest in the magnetosheath \citep{robertson_cravens03_b} and often occurs at times when the solar wind proton flux is high \citep{carter_sembay08, kuntz_snowden08}.  Heliospheric SWCX tends to be stronger near the ecliptic plane \citep{robertson_cravens03_a, koutroumpa_etal06} and varies with the overall solar cycle (11 yr).  Fortunately, techniques to reduce SWCX emission to \ion{O}{7} and \ion{O}{8} emission line measurements exist \citep{carter_sembay08, carter_etal11}.  

\begin{figure}
\centering
\includegraphics[width = 1.0\textwidth, keepaspectratio=true]{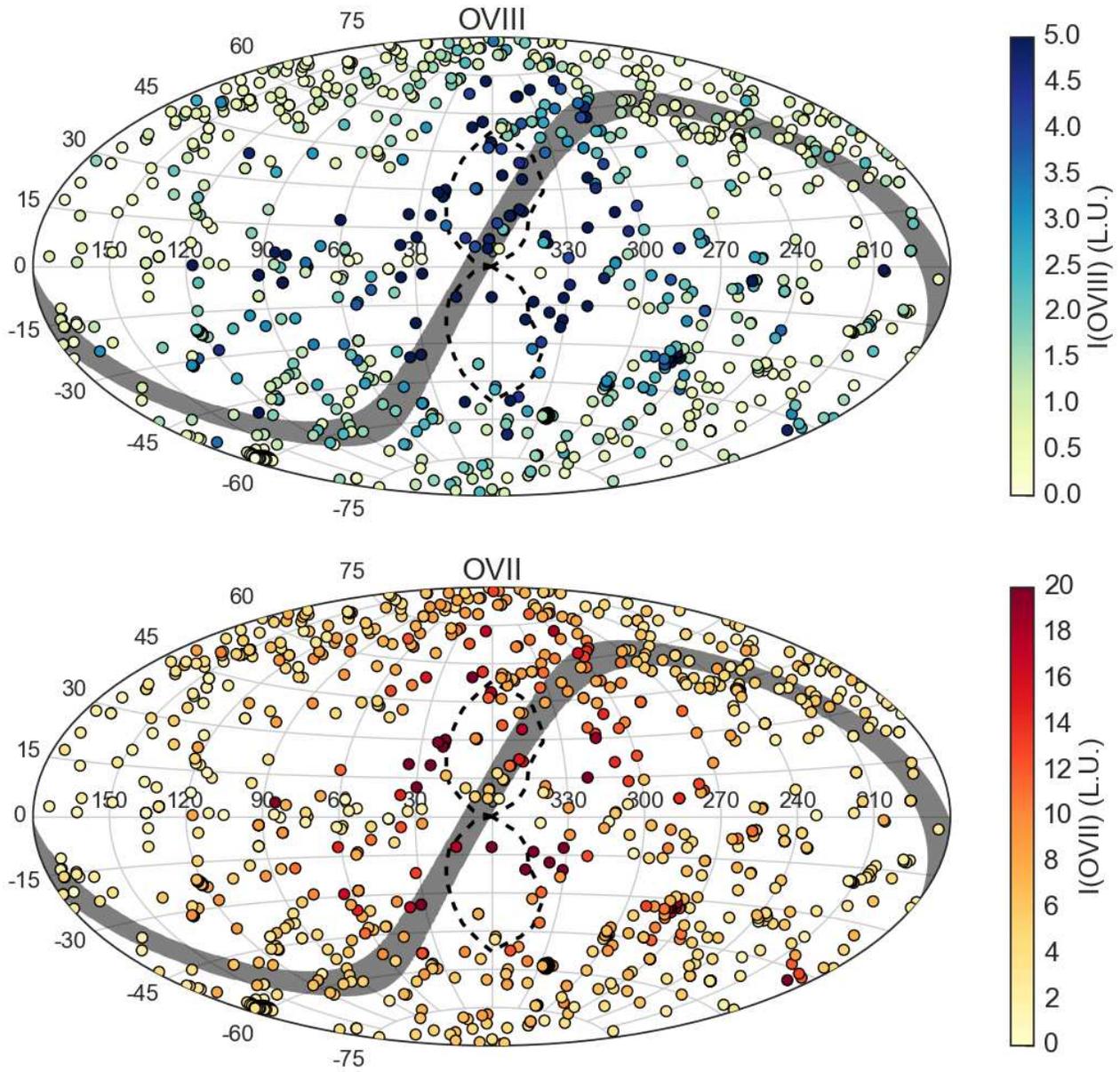}
\caption{Flux-filtered sample of \ion{O}{8} (top) and \ion{O}{7} (bottom) emission line strengths across the sky from \cite{hs12}.  The dashed line represents the observed gamma-ray emission from the Fermi bubbles \citep{su_etal10} and the shaded gray strip represents 5$\arcdeg$ above and below the ecliptic plane.
}
\label{figure.map_filter}
\end{figure}

\cite{hs12} address several SWCX filtering techniques in detail, but their primary approach utilizes solar wind proton flux data from OMNIWeb \footnote{\url{http://omniweb.gsfc.nasa.gov/}}.  This database includes solar wind proton flux data from numerous satellites, including the \textit{Advanced Composition Explorer} and \textit{Wind}.  They reduce SWCX contamination to the emission lines by removing portions of the \textit{XMM-Newton} data taken when the solar wind proton flux was greater than $2 \times 10^{8}$ cm$^{-2}$ s$^{-1}$.  This filtering procedure caused the number of usable observations to fall from 2611 to 1435 due to some of the observations falling below the 5 ks observing time threshold mentioned above.  The authors presented both their full sample of emission lines and this ``flux-filtered'' sample of emission lines that are less contaminated by SWCX emission.  Figure~\ref{figure.map_filter} shows the distribution of the flux-filtered sample on the sky and the corresponding emission line strengths.

\subsection{Emission Line Measurements}
\label{subsection.line_measurements}

In this section, we outline the emission line measurement procedure from \cite{hs12}.  This includes fitting for all emission sources in a typical SXRB spectrum, including the emission lines of interest, the continuum of the LB and hot halo, the extragalactic background (EPL), the instrumental lines in the 0.4 - 10.0 keV range, and any residual soft proton contamination present after filtering out the quiescent particle background (QPB).  This entire measurement procedure used XSPEC \footnote{\url{http://heasarc.gsfc.nasa.gov/docs/xanadu/xspec/}} version 12.7.0.  

The authors fitted each observation in the 0.4 - 10.0 keV range with a multicomponent spectral model discussed above.  The \ion{O}{7} and \ion{O}{8} emission line model consisted of two $\delta$-functions (or Gaussians with widths fixed to zero) with the \ion{O}{7} energy centroid left as a free parameter and the \ion{O}{8} energy centroid fixed at 0.6536 keV \citep[from APEC;][]{smith_etal01}.  Note that this line measurement method includes the total line emission from all sources in the extracted spectrum (hot halo, LB, and any residual contamination).  The Galactic continuum included an absorbed APEC thermal plasma model \citep{smith_etal01} with the oxygen K$\alpha$ emission disabled \citep{lei_etal09}.  The EPL was modeled as an absorbed power law with a photon index of 1.46 \citep{chen_etal97}.  Both the APEC and EPL components included attenuation by absorbing H\,{\sc i} columns from the LAB survey \citep{kalberla_etal05} using the XSPEC \texttt{phabs} model \citep{bc_mccammon92, yan_etal98}.  The APEC and \texttt{phabs} components assumed \cite{anders_grevesse89} abundances.  The model included two Gaussians for the Al-K and Si-K instrumental lines in this energy range \citep{kuntz_snowden08}.  The final model component included a power law (not folded through the instrumental response) to account for residual soft proton contamination from the QPB \citep{snowden_kuntz11}.  

\cite{hs12} provide a detailed discussion of the assumptions made in the above model and how these assumptions relate to the uncertainties presented in their results.  Here we highlight those that affect the uncertainties of the emission line measurements.  The statistical errors of the emission line measurements come from the standard XSPEC \texttt{error} command.  The authors also provide a systematic uncertainty estimate based on the type of thermal plasma model used and variation in the EPL normalization parameter.  The former compares the original line intensity measurements assuming an APEC thermal plasma model with measurements assuming a \normalsize{M}\footnotesize{E}\normalsize{K}\footnotesize{A}\normalsize{L} \citep{mewe_etal95} or \cite{raymond_smith77} model \citep[see Equation (1) in][]{hs12}.  The latter accounts for sightline-to-sightline variation in the EPL normalization parameter due to variable soft proton contamination and/or unresolved sources with $F_X^{0.5-2.0}$ \textless $5 \times 10^{-14}$ erg cm$^{-2}$ s$^{-1}$ \citep{moretti_etal03}.  The total systematic error for each observation is the combination of these two estimates in quadrature.

The final filtering procedure discussed in \cite{hs12} applies a more restrictive constraint on residual soft proton contamination to the emission line measurements.  The authors introduced the ratio between the total 2 - 5 keV band flux ($F_{total}^{2-5}$) and the EPL flux in the same energy band ($F_{exgal}^{2-5}$) as a quantitative measure of residual soft proton contamination to the count rates \citep{hs10}.  Any observations where this ratio was greater than 2.7 were rejected from the sample.  This decreased the number of useful observations from 2611 to 1868 for their full sample and from 1435 to 1003 for their flux-filtered sample.

\subsection{Additional Observation Screening}
\label{subsection.our_screening}

We apply additional screening methods to the data in order to produce a sample that is most sensitive to the Milky Way's hot halo emission.  These are spatial screening methods where we discard observations that are located near possible contaminates to the \ion{O}{7} and \ion{O}{8} emission lines.  These include bright X-ray sources in the sky and any Galactic features that show evidence for enhanced soft X-ray emission (see Table~\ref{table.screening_catalogs} for a summary).  Although \cite{hs12} account for a range of point source removal methods in their data reduction procedure (see Section ~\ref{subsection.data_filtering}), this additional screening is designed to provide a cleaner sample of Milky Way hot halo emission.  

Our automated screening discards any observations within 0.5$\arcdeg$ (or within the \textit{XMM-Newton} field of view) of sources that could produce soft X-rays, thus complicating the \ion{O}{7} or \ion{O}{8} emission line measurement.  We utilize several all-sky catalogs that cover a range of astrophysical objects to generate a potential contaminating source list.  The final contaminating source list is a subset of the objects in these catalogs based on either direct or inferred X-ray brightness cuts.  For example, for \textit{ROSAT} Bright Source Catalog objects we include sources with fluxes \textgreater 1 counts s$^{-1}$, but for Principal Galaxy Catalog objects, which do not have observed X-ray fluxes, we include sources with sizes \textgreater 10$\arcmin$.  This does not create a completely uniform list of different types of X-ray-emitting objects and their fluxes, but it does provide a general list of objects that could contaminate the \ion{O}{7} or \ion{O}{8} emission line measurement.  

We also discard observations by hand that are known Galactic X-ray features and that are not discarded by our automated screening procedure outlined above.  We remove any observations within $\leq 10 \arcdeg$ of the Galactic plane to reduce emission from SNe in the Milky Way's disk \citep{norman_ikeuchi89, joung_maclow06, hill_etal12}.  This region also contains the largest H\,{\sc i} columns that attenuate hot halo line emission, thus making the de-absorbed emission line intensities more uncertain.  We also remove observations within $|l| \leq 22 \arcdeg$ and $|b| \leq 55 \arcdeg$, or near the Fermi Bubbles \citep[e.g.,][]{su_etal10}.  It is unclear how the bubbles have impacted the Milky Way's hot gas halo, but there are signatures of the bubbles in X-ray spectra \citep{bh_cohen03, kataoka_etal13}.  The observations that pass through these regions will be part of a separate study.  Finally, we remove a cluster of observations near the Large Magellanic Cloud and Small Magellanic Cloud.

\begin{figure}
\centering
\includegraphics[width = 1.0\textwidth, keepaspectratio=true]{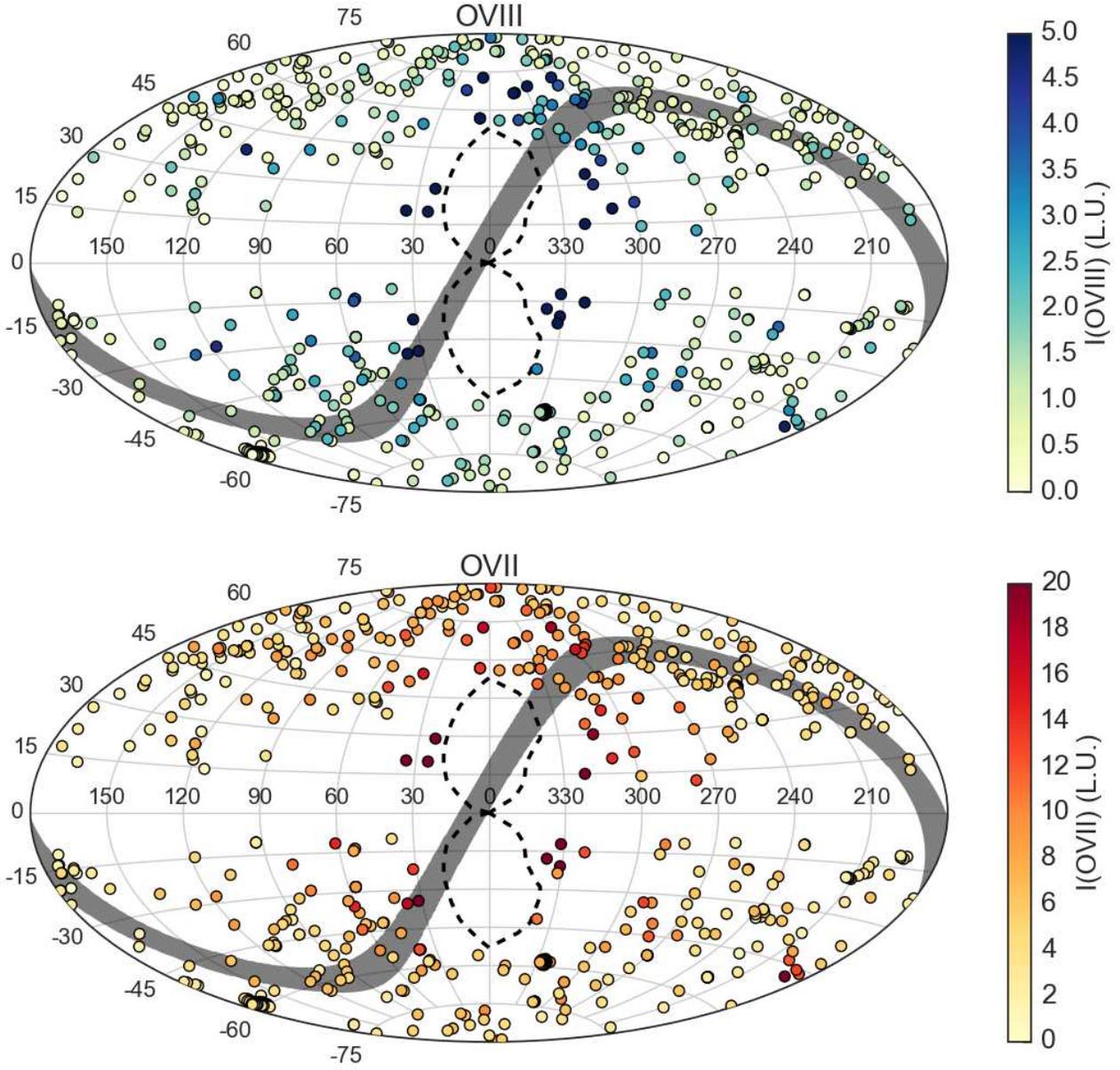}
\caption{Same as Figure~\ref{figure.map_filter} but with our additional screening criteria applied to the observations (see Section ~\ref{subsection.our_screening}).  Note the emission line strengths tend to increase from Galactic anticenter toward the Galactic center.  These observations serve as our sample in our model fitting procedure.  
}
\label{figure.map_spacecut}
\end{figure}

After our additional screening criteria discussed above, we are left with 649 of the 1003 observations from the flux-filtered sample.  Figure~\ref{figure.map_spacecut} shows the distribution of our screened sample on the sky with the corresponding \ion{O}{7} and \ion{O}{8} emission line strengths.  Other than the Galactic plane and Fermi Bubble regions, our screened sample has similar sky coverage compared to the original all-sky samples.  This sub-sample of observations should contain minimal emission from sources other than the LB and Milky Way hot halo and serves as our working sample of observations in our model fitting procedure.

\section{Model Fitting}
\label{section.model_fitting}

Here we describe how we compare the emission line observations from \cite{hs12} with simulated line intensities from a parametric density model for the Milky Way's hot gas halo.  The goal of this work is to explore our model parameter space and find the set of parameters that is most consistent with the observations.  This way we constrain the density profile of the Milky Way's hot gas halo with the same approach as previous studies on \ion{O}{7} absorption lines \citep{miller_bregman13}.

\subsection{Halo Density Model}
\label{subsection.halo_model}

We consider various forms of a spherical $\beta$-model to represent the Milky Way's hot gas halo density profile.  We choose this model because of its success in fitting X-ray surface brightness profiles around early- \citep{forman_etal85, osullivan_etal03} and late-type \citep{anderson_bregman11, dai_etal12} galaxies and to be consistent with previous work on the Milky Way's hot halo \citep{miller_bregman13}.  The $\beta$-model is defined as

	\begin{equation}
	 \label{eq.beta_model}
	 n(r) = {n_o}({1 + ({r}/{r_c})^2})^{-{3\beta}/{2}},
	\end{equation}

\noindent where $n_\circ$ is the central density, $r_c$ is the core radius, and $\beta$ defines the slope of the profile at large radii.  Typical values for $\beta$ range between 0.4 and 1.0 while typical core radii are $\lesssim$ 5 kpc.  

The typical observed values for the core radii parameter combined with our removal of observations near the Galactic center limit our ability to simultaneously constrain all three parameters of the $\beta$-model.  Out of the 649 pointings in our sample discussed above, only 4 pass within 5 kpc of the Galactic center.  Following the work of \cite{miller_bregman13}, we adopt a modified $\beta$-model in the limit where $r \gg r_c$:

	\begin{equation}
	 \label{eq.beta_model_approx}
	 n(r) \approx \frac{n_or_c^{3\beta}}{r^{3\beta}}.
	\end{equation}

\noindent  This modified density profile is essentially a power law with a normalization of $n_\circ r_c^{3\beta}$ and slope of -3$\beta$.  We use this two-dimensional hot gas halo density profile for the majority of our analysis.  

We also consider a flattened hot gas halo density profile in our analysis by modifying the spherical $\beta$-model.  A flattened density profile would be a natural consequence of the Milky Way's rotation.  If the hot gas halo traces the dark matter (DM) distribution and the DM distribution is flattened due to rotation, one expects some flattening of the hot gas density profile \citep{stewart_etal13}.  The flattened $\beta$-model becomes

	\begin{equation}
	 n(R,z) = {n_o}({1 + ({R}/{R_c})^2 + ({z}/{z_c})^2})^{-{3\beta}/{2}},
	\end{equation}

\noindent where $R$ and $z$ are the radius in the plane of the Milky Way's disk and height off the Galactic plane, respectively, and $R_c$ and $z_c$ are the core radii in those directions.  We infer a flattening of the density profile based on the ratio between $R_c$ and $z_c$.

\subsection{LB/SWCX Model}
\label{subsection.LB_model}

We have to model all other known sources of \ion{O}{7} and \ion{O}{8} line emission (e.g., the LB and residual SWCX) simultaneously with our hot halo density model.  This is because the observed emission lines contain the total emission from all these sources and because the sources may produce emission line strengths comparable to one another.  However, we emphasize that the goal of this work is to understand properties of the Milky Way's hot gas halo, not the LB or SWCX.  We thus develop a parametric model for the LB (similar to the hot gas halo), but we do not claim this parameterization to be the correct physical interpretation for the LB.  This still parameterizes the \textit{emission} properties of the LB, which is necessary for correctly interpreting the hot gas halo emission.  

We model any emission not due to the Milky Way's hot halo as a volume-filled, constant-density LB.  The density and temperature along every sightline is assumed to be the same while the path length of the LB ranges between $\approx$100-300 pc depending on the direction of the observation.  These path lengths come from \ion{Na}{1} absorption line measurements of 1005 nearby ($\lesssim$350 pc) stars \citep{lallement_etal03}.  Figures 4-6 in \cite{lallement_etal03} show contours separating local ionized and neutral gas, or equivalently defining an LB edge.  The contours show a variety of substructure associated with the LB, although there is a general shape defining the ionized gas region.  In our analysis, we parameterize the LB edges with a geometrical model to approximately match the contours discussed above.  This implies that the LB emission varies by a factor of $\sim$3, depending on the observed direction.  We note this parameterization may not accurately represent the physical properties of the LB, specifically the constant-density and volume-filled assumptions.  There are reports indicating that the LB is better described as a ``cavity'' where the emission comes from a wall of material with little emission from diffuse, volume-filling material \citep[e.g.,][]{welsh_shelton09}.  However, other works have modeled the LB X-ray emission properties assuming that the bubble is filled with a diffuse plasma along the line of sight \citep{smith_etal07}.  Differentiating between these scenarios is beyond the scope of this work since this simple parameterization still characterizes the X-ray emission of the LB with some physical motivation.  

\subsection{Temperature Assumptions}
\label{subsection.temperature_assumptions}

All \ion{O}{7} and \ion{O}{8} line emissivities come from AtomDB version 2.0.2 \citep{foster_etal12}.  The line emissivities assume an APEC thermal plasma in collisional ionization equilibrium (CIE) at a given temperature \citep{smith_etal01}.  We initially assume that the halo and LB plasmas are isothermal with temperatures of log($T_{halo}$) = 6.3 and log($T_{LB}$) = 6.1, resulting in \ion{O}{7} and \ion{O}{8} line emissivities (in units of photons cm$^3$ s$^{-1}$) of $\epsilon_{\text{\ion{O}{7}}}$ (halo) = $6.05 \times 10^{-15}$, $\epsilon_{\text{\ion{O}{8}}}$ (halo) = $1.45 \times 10^{-15}$, $\epsilon_{\text{\ion{O}{7}}}$ (LB) = $1.94 \times 10^{-15}$, and $\epsilon_{\text{\ion{O}{8}}}$ (LB) = $2.67 \times 10^{-17}$.  Note that the \ion{O}{7} line emissivities quoted here include the resonance, forbidden, and intercombination lines since the observed emission lines are unresolved.

We assume that the hot gas halo is a constant-temperature plasma in CIE with fixed log($T_{halo}$) = 6.3.  The strongest observational evidence for the hot gas halo being isothermal comes from modeling the 0.5 - 2.0 keV spectra of the SXRB for a large number of sightlines.  \cite{hs13} analyzed 110 observations of the SXRB based on a subset of the observations from \cite{hs12}.  Their spectral fitting routine was similar to the line measurement procedure above, except they fit the spectra with thermal plasma models for the hot gas halo instead of measuring the individual \ion{O}{7} and \ion{O}{8} line intensities.  This resulted in hot gas halo emission measures and temperatures for a subsample of the \cite{hs12} observations.  The primary benefit of this approach is the ability to fit for the hot gas halo emission measure and temperature separately from the LB, as opposed to the emission lines that include all emission sources.  The drawback to this spectral fitting technique is that one requires more counts to constrain the plasma properties as opposed to just the emission line surface brightnesses, thus requiring more exposure time per observation.  We note that \cite{hs13} discuss an in-preparation study indicating the emission measure and line intensity observations are generally consistent with each other.  Furthermore, the sample analyzed in \cite{hs13} is subjected to stronger temporal and spatial screening criteria than the \cite{hs12} sample that limit the sample's sky coverage.  Thus, the significant increase in sample size and sky coverage of the emission line measurements compared to the emission measure/temperature measurements imply the former are better suited for this study.  This study still found that the hot halo temperature showed little variation with a median of $2.22 \times 10^6$ K with an interquartile range of $0.63 \times 10^6$ K (we discuss their emission measure results in Section ~\ref{subsection.previous_work_obs}).  This is also consistent with previous temperature estimates of the hot gas halo from its emission properties \citep{mccammon_etal02, yoshino_etal09}, thus validating this assumption.  

As discussed above, we also assume that the LB is a constant-temperature plasma in CIE with fixed log($T_{LB}$) = 6.1.  Similar to our constant-density parameterization, this assumption is also debated in the literature.  \cite{smith_etal07} conducted a shadowing experiment toward the nearby molecular cloud MBM12 to constrain the \ion{O}{7} and \ion{O}{8} line emission due the LB.  They found that the ratio of the emission line strengths implied an LB plasma temperature of $1.2 \times 10^6$ K.  However, recent work also implies that the LB may not be in CIE and that non-equilibrium ionization effects can change the interpretation of LB observations \citep{deavillez_etal13}.  Although departures from CIE are crucial for understanding the physical properties of the LB, the work by \cite{smith_etal07} motivates our temperature assumption given our model parameterization.  

\subsection{Optical Depth Considerations}
\label{subsection.tau_considerations}

There has been growing evidence for optical depth effects associated with the Milky Way's hot halo plasma.  This has been a difficult effect to quantify since both the emission and absorption lines are unresolved, making direct measurements of the linewidths impossible.  The best evidence for optical depth effects in the plasma comes from weak detections of \ion{O}{7} $K \beta$ absorption lines in several QSO spectra \citep{bregman_ld07, williams_etal05, gupta_etal12}.  These works imply optical depths ranging between $\approx$ 0.1 - 2.0 depending on the Doppler width assumed for the lines, whether one is observing \ion{O}{7} or \ion{O}{8}, and which direction one observes \citep{gupta_etal12, miller_bregman13}.

These optical depth estimates mean assuming that the plasma is optically thin (which is often an implicit assumption) may be incorrect.  To estimate the optical depth corrections, we calculate line intensities both in the optically thin limit and assuming a Doppler width of $b$ = 150 km s$^{-1}$.  This Doppler width is characteristic of the hydrogen sound speed at this temperature and is consistent with simulations of halo gas \citep{fukugita_peebles06, cen12}.  The Doppler width is necessary for calculating the absorption cross section of the line transitions (see below).  Since these effects are difficult to quantify, we present results both by calculating emission lines in the optically thin limit and by accounting for optical depth corrections.  

\subsection{Line Intensity Calculation}
\label{subsection.intensity_calculation}

The model line intensity along each line of sight involves a unique calculation given the geometry of the halo, the LB, and our proximity to the Galactic center.  For every ($l,b$) coordinate, there is a unique coordinate transformation between that line of sight distance and galactocentric radius.  This coordinate tranformation is summarized by the following equations:

	\begin{equation}
		R^2 = R_o^2 + s^2cos(b)^2 - 2sR_ocos(b)cos(l)
	\end{equation}
	\begin{equation}
		z^2 = s^2sin(b)^2
	\end{equation}
	\begin{equation}
		r^2 = R^2 + z^2.
	\end{equation}

\noindent In these equations, $R$ and $z$ are respectively the distance from the Galactic center in the plane of the Milky Way's disk and height off the Galactic plane, $r$ is the galactocentric radius, $R_o$ is the between the Sun and the Galactic center (we assume $R_o$ = 8.5 kpc), and $s$ is the line of sight distance.  These are the equations we use to evaluate the halo emission contribution along every line of sight.  

The simplest line intensity calculation we make assumes an optically thin hot halo plasma.  The transfer equation under this assumption is

	\begin{equation}
	 \label{eq.transfer_thin}
	 \frac{dI_{thin}}{ds} = j(s) = \frac{n_{halo}(s)^2 \epsilon(T_{halo})}{4 \pi},
	\end{equation}
	
\noindent where $n_{halo}(s)$ is the halo density profile along a given line of sight and $\epsilon(T_{halo})$ is the halo line emissivity discussed in Section ~\ref{subsection.temperature_assumptions}.  This leads to the integral form of our optically thin emission line calculation:

	\begin{equation}
	 \label{eq.intensity_thin}
	 I_{thin} (l,b) = \frac{1}{4 \pi} \int n_{halo}(s)^2 \epsilon(T_{halo}) ds,
	\end{equation}
	
\noindent where we integrate the halo density profile out to the Milky Way's virial radius ($\approx$250 kpc).  

We also calculate line intensities while estimating optical depth corrections in the plasma (see Section ~\ref{subsection.tau_considerations}).  This calculation is different than the optically thin calculation by accounting for self absorption/scattering of photons by the plasma and scattering of photons into the line of sight.  This changes the transfer equation in the following way:

	\begin{equation}
	 \label{eq.transfer_optical_depth}
	 \frac{dI_{\tau}}{ds} = j(s) - \kappa (s) I(s) + \kappa (s) J(s),
	\end{equation}
	
\noindent where $j(s)$ is the same as in Equation (\ref{eq.transfer_thin}).  The absorption coefficient ($\kappa (s)$) quantifies the interaction between photons and ions and is defined as
	
	\begin{equation}
    \kappa (s) = n_{halo} (s) A_O X_{ion} \times \sigma = n_{halo} (s) A_O X_{ion} \times .015 f \lambda b^{-1},
	\end{equation}
	
\noindent where $A_O$ is the oxygen abundance relative to hydrogen, $X_{ion}$ is the ion fraction of the absorbers, $\sigma$ is the absorption cross section, .015 is a constant with units of cm$^2$ s$^{-1}$, $f$ is the oscillator strength of the transition, $\lambda$ is the transition wavelength in centimeters, and $b$ is the Doppler width of the lines in cm s$^{-1}$.  

\cite{anders_grevesse89} abundances are the most commonly cited solar abundance values, but more current estimates advocate a solar oxygen abundance between 35\% and 50\% lower than the \cite{anders_grevesse89} value \citep{holweger_01, asplund_etal05}.  Here, we adopt the \cite{anders_grevesse89} oxygen abundance (log($N_O$)=8.93) to be consistent with previous work and with APEC metal abundances.  We also assume $X_{ion}$ of 0.5 for both \ion{O}{7} and \ion{O}{8} \citep{sutherland_dopita93} and a Doppler width of 150 km s$^{-1}$.

The positive scattering term in Equation (\ref{eq.transfer_optical_depth}) accounts for single-scattered photons into the line of sight and depends on the mean intensity at every point along the line of sight ($J(s)$).  The mean intensity follows the formal definition:

	\begin{equation}
    J(r) = \frac{1}{4 \pi} \int_{4\pi} I(r, \Omega) d\Omega.
	\end{equation}

\noindent We introduce the standard definition of optical depth here, $d\tau = \kappa (s) ds$, to represent Equation (\ref{eq.transfer_optical_depth}) in a simpler way:

	\begin{equation}
	 \frac{dI_{\tau}}{d\tau} = \frac{j(s)}{\kappa (s)} - I(s) + J(s) = S(s) - I(s),
	\end{equation}
	
\noindent where $S(s)$ is the source function along the line of sight, defined as $j(s) / \kappa (s) + J(s)$.  This differential equation can be solved to represent our line intensity calculation accounting for optical depth effects:
	
	\begin{equation}
	 I_{halo, \tau} (l,b) = \int e^{-(\tau_\circ - \tau)} S(s) d\tau,
	\end{equation}
	
\noindent where $\tau_\circ = \int \kappa (s) ds$ is the total optical depth for that line of sight.

These optical depth corrections are an approximation in the single-scattered photon case.  The expression for $J(r)$ assumes we know the true mean intensity (or photon density) at every location in the halo for a given set of model parameters.  Thus, for each model parameter set, we calculate $J(r)$ once to estimate the scattering contribution to the emission lines.  This is only an approximation because $J(r)$ depends on $I$ from all directions, which inherently depends on $J(r)$.  We discuss this approximation in more detail in Section ~\ref{subsection.tau_effects}.

The LB line intensity calculation is more straightforward than our halo calculation due our constant density and temperature assumptions.  We treat the LB as an optically thin plasma, thus making the line intensity calculation the limiting case of Equation (\ref{eq.intensity_thin}) for a constant-density plasma.  The calculation simplifies to

	\begin{equation}
	 I_{LB} (l,b) = \frac{n_{LB}^2 L(l,b) \epsilon(T_{LB})}{4 \pi},
	\end{equation}

\noindent where $n_{LB}$ is our LB density parameter, $L(l,b)$ is the LB path length inferred from \cite{lallement_etal03}, and $\epsilon(T_{LB})$ is the LB line emissivity discussed in Section ~\ref{subsection.temperature_assumptions}.

We finally add our halo and line intensities together to make a total model line calculation defined as

	\begin{equation}
	 I_{total} (l,b) = I_{LB} (l,b) + I_{halo} (l,b) e^{-\tau_{HI}},
	\end{equation}

\noindent where the $e^{-\tau_{HI}}$ term accounts for \ion{H}{1} absorption in the Galactic disk.  Here we define $\tau_{HI} = \sigma_{HI} N_{HI}$ where $\sigma_{HI}$ is the \ion{H}{1} absorption cross section \citep{bc_mccammon92, yan_etal98} and $N_{HI}$ is the column density for a given line of sight from the LAB survey \citep{kalberla_etal05}.  This is a necessary step since the observed line intensity measurements account for the total line emission due to all emission sources along the line of sight.  Note we assume the LB is the only plasma component between the Sun and $L(l,b)$, and the halo line intensity integration goes from $L(l,b)$ to $R_{vir}$ along a given line of sight.  We still attenuate all our model halo emission with the observed \ion{H}{1} column densities since starting our halo integration at $\sim$100-300 pc compared with distances securely beyond the Milky Way's \ion{H}{1} disk ($\sim$1 kpc) results in $\lesssim$5\% differences in the halo line emission.  In this way, we calculate a model line intensity along any location in the sky given a set of halo and LB density parameters.  

\subsection{Fitting Procedure}
\label{subsection.fitting_procedure}

The purpose of our model fitting procedure is to find the parameters for our emission model that best reproduce the observed line intensities.  Quantitatively, our goal is to minimize the $\chi^2$, or maximize the likelihood $L \propto$ exp$(-.5\chi^2)$ in our case, between our model and the observations.  We utilize the Markov Chain Monte Marlo (MCMC) package \texttt{emcee} \citep{foreman_mackey_etal13} to explore our model parameter space (note we assume uniform prior distributions for all MCMC runs unless otherwise noted).  This package is a Python implementation of Goodman \& Weare's Affine Invariant MCMC Ensemble sampler designed for parameter estimation \citep{goodman_weare10}.  We define the input ln($L$) as $-.5\chi^2$, generate a random set of starting points for our parameters, and allow the code to explore the parameter space to maximize the likelihood.  

We examine the marginalized posterior probability distributions of our model parameters to determine the parameter set that best reproduces our observed line intensities.  The \texttt{emcee} package conveniently outputs these distributions for each sampler run on a set of line observations and for a given plasma type (optically thin or with optical depth corrections).  The binned distributions are then fit with Gaussian functions with the centroid being the optimal parameter and the $\sigma$ parameter being the uncertainty of the optimal parameter.  In this way, we constrain the model parameter set that is most consistent with the observations.

\section{Results}
\label{section.results}

Here we present our results based on our model fitting procedure discussed above.  We spend most of our analysis fitting the \ion{O}{7} and \ion{O}{8} emission line observations separately rather than both measurements simultaneously.  This is the better approach since the typical signal-to-noise ratio (S/N) of the \ion{O}{7} measurements (mean = 4.9) are larger than the \ion{O}{8} measurements (mean = 1.3) in our sample.  We also present results with and without optical depth effects present in the line intensity calculation.  A summary of our results can be found in Table~\ref{table.mcmc_results}.  Note that throughout the rest of the paper, our quoted $\chi^2$ and $\chi_{red}^2$ values are from the best-fit values in Table~\ref{table.mcmc_results}.  We use these metrics to quantify the consistency between our models and the observations.  

\subsection{Analyzing \ion{O}{8} Line Emission}
\label{subsection.oviii_analysis}

\begin{figure}
\centering
\includegraphics[width = 1.0\textwidth, keepaspectratio=true]{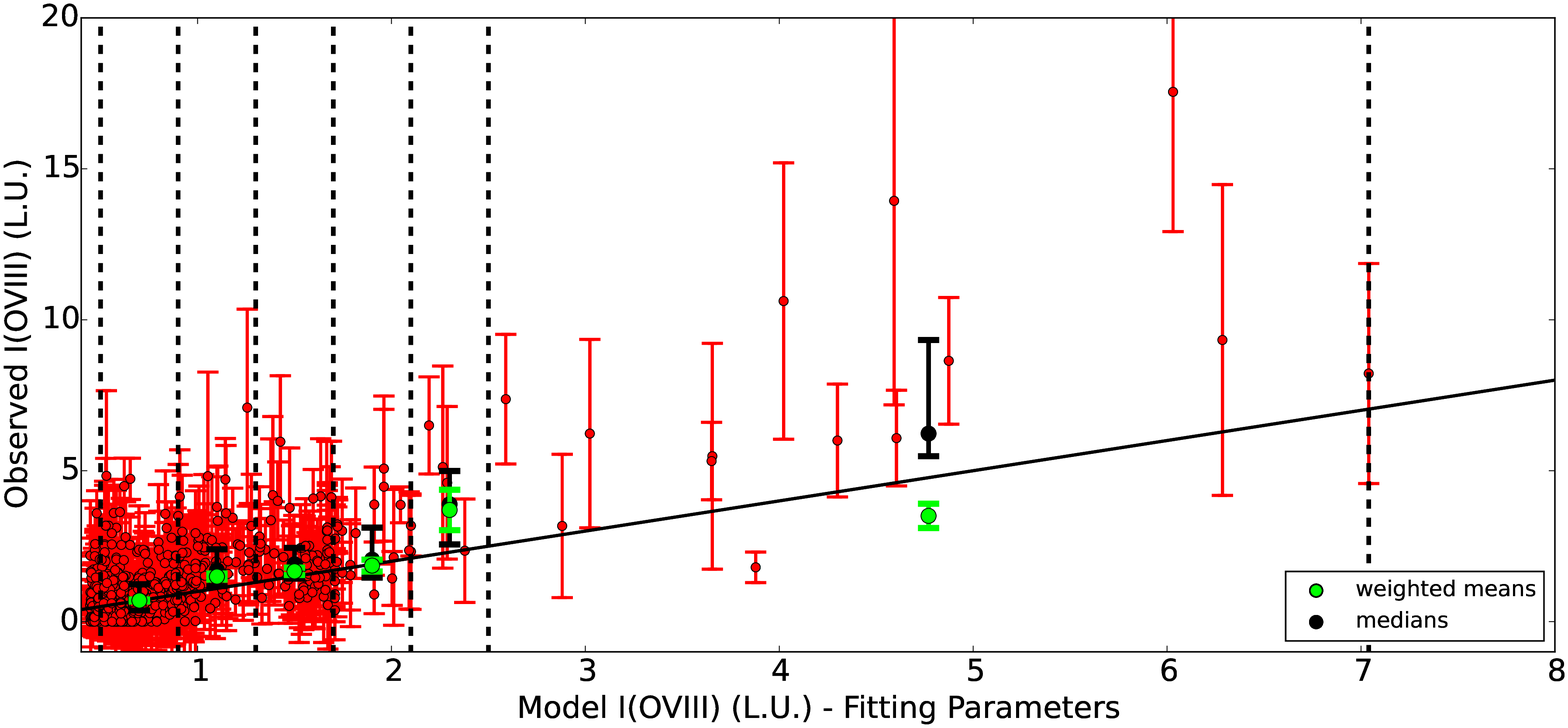}
\caption{Observed \ion{O}{8} emission line values compared with our best-fit model values assuming an optically thin plasma.  The error bars on the observations are the addition of statistical and systematic uncertainties in quadrature.  We bin the data and show the medians with first through third quartile regions (black points) along with the weighted means (green points).  The vertical black dotted lines are the bin edges used while the black solid line represents the one-to-one line.  The binned data indicate our best-fit model reproduces the data.  }
\label{figure.obs_vs_mod}
\end{figure}

The \ion{O}{8} emission line observations are fit very well with our parametric model ($\chi_{red}^2$ (dof) = 1.08 (644)) regardless if we assume the plasma is optically thin or apply our optical depth corrections.  The $\beta$ parameter ranges between 0.50 and 0.54 depending on the type of plasma we assume, corresponding to a $\sim r^{-3/2}$ density profile.  Figure~\ref{figure.obs_vs_mod} shows the relationship between the observed \ion{O}{8} and our best-fit model intensities, assuming that the plasma is optically thin.  This includes weighted means and medians (with interquartile regions) for several model intensity bins to show that our best-fit model reproduces the \ion{O}{8} observations.  Note that we use smaller bin sizes for $I \lesssim$ 2 L.U. (where the unit L.U. = photons cm$^{-2}$ s$^{-1}$ sr$^{-1}$) to show that we reproduce the general properties of the \ion{O}{8} observations with our best-fit model.

The quality of our fit to the \ion{O}{8} observations significantly improves with the exclusion of one outlier observation (\textit{XMM-Newton} Observation ID 0200730201, ($l,b$) = 327.59$\arcdeg$, +68.92$\arcdeg$).  This observation  has an abnormally large line strength and S/N ($I_{obs}$ = 8.69 L.U., S/N with total uncertainty = 10.84) compared to the expected model value in this direction ($I_{mod}$ = 1.18 L.U.), making it a $\approx 9 \sigma$ outlier.  We ran our model fitting procedure on our \ion{O}{8} emission line sample with and without this observation included and found that the best-fit parameters did not change, but the $\chi_{red}^2$ (dof) dropped significantly from 1.21 (645) to 1.08 (644).  This observation was subjected to the standard flux-filtering procedure discussed above, so we do not expect SWCX contamination to be the cause of this peculiarity.  It is, however, located at the tip of the north polar spur, a known region of enhanced X-ray emission thought to be caused by a nearby SN remnant \citep{miller_etal08}.  We intend to examine this region in more detail in the future, but the quality of our fit results and density model constraints with its removal from the sample indicates that the \ion{O}{8} observations are well described by our model.  

We allow the hot gas halo size to vary as a free parameter to estimate the minimum size halo necessary to still be consistent with the \ion{O}{8} observations.  Initially, we calculate line intensities assuming a halo size of 250 kpc, but our line intensity calculation becomes less sensitive at larger galactocentric radii due to the decrease in density.  In other words, our model line intensity values do not change much once we integrate past a line of sight distance of $\approx$50 kpc.  When we allow the halo size to vary, we find nearly identical posterior probability distributions for the original model parameters.  The distribution for the halo size is not a well-defined Gaussian like the other parameters, but instead is a flat distribution between 15 and 250 kpc.  Thus, we find a minimum halo size of $\approx$15 kpc.  This size scale is consistent with a large, extended gas distribution, rather than a compact disk morphology.  \cite{miller_bregman13} also found similar minimum halo sizes of 32 kpc and 18 kpc at the 95\% and 99\% confidence levels from a similar analysis of \ion{O}{7} absorption lines.  

Similar to the work by \cite{miller_bregman13}, we had trouble fitting a flattened density model to the observations.  This is likely due to the inferred vales for $r_c, R_c,$ and $z_c$ ($\lesssim$1 kpc) and our lack of observations near the Galactic center.  The only way we can constrain $R_c$ and $z_c$ for the flattened model is by limiting the explored parameter space of the other free parameters in our model.  We employ boundaries of 0.01 - 0.50 cm$^{-3}$ for $n_o$, 0.3-0.8 for $\beta$, and 0.0 - 0.1 for $n_{LB}$ in our MCMC analysis while $R_c$ and $z_c$ are left to explore their full parameter space.  When we set these boundaries and assume an optically thin plasma, we find a best-fit model with $R_c$ = 0.23 $\pm$ 0.09 kpc and $z_c$ = 0.27 $\pm$ 0.10 kpc ($\chi_{red}^2$ (dof) = 1.08 (644)).  The orthogonal core radii are consistent with each other based on their 1$\sigma$ uncertainties, indicating that a flattened density profile is not an improvement over our spherical density model.  

\begin{figure}
\centering
\includegraphics[width = 1.0\textwidth, keepaspectratio=true]{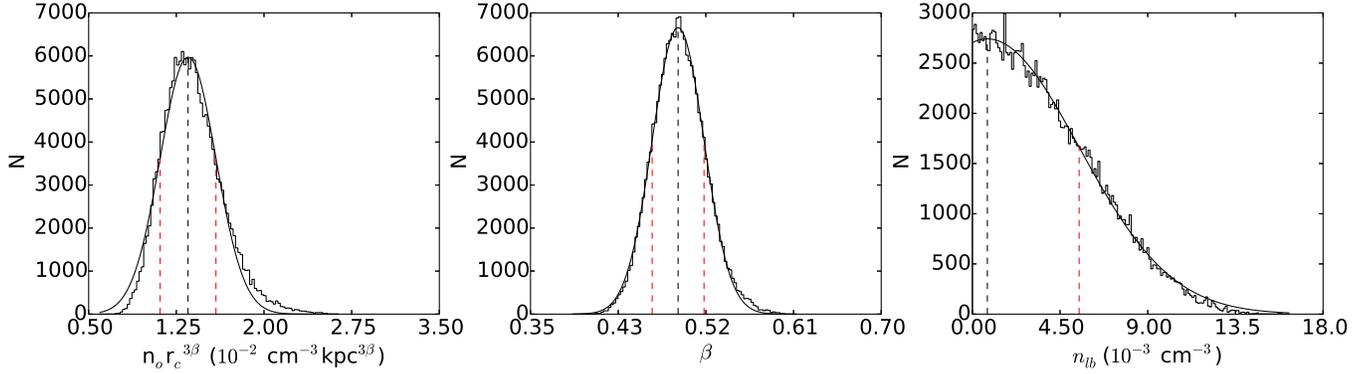}
\caption{Marginalized posterior probability distributions of our model parameters while fitting the \ion{O}{8} observations with an optically thin plasma.  The smooth black curve is a fitted Gaussian function to the distributions while the dashed lines represent the Gaussian centroid (black) and 1$\sigma$ (red) parameters.  These dashed lines in each plot represent the best-fit parameter values and uncertainties quoted in Table~\ref{table.mcmc_results}.  
}
\label{figure.chains}
\end{figure}

\begin{figure}
\centering
\includegraphics[width = 1.0\textwidth, keepaspectratio=true]{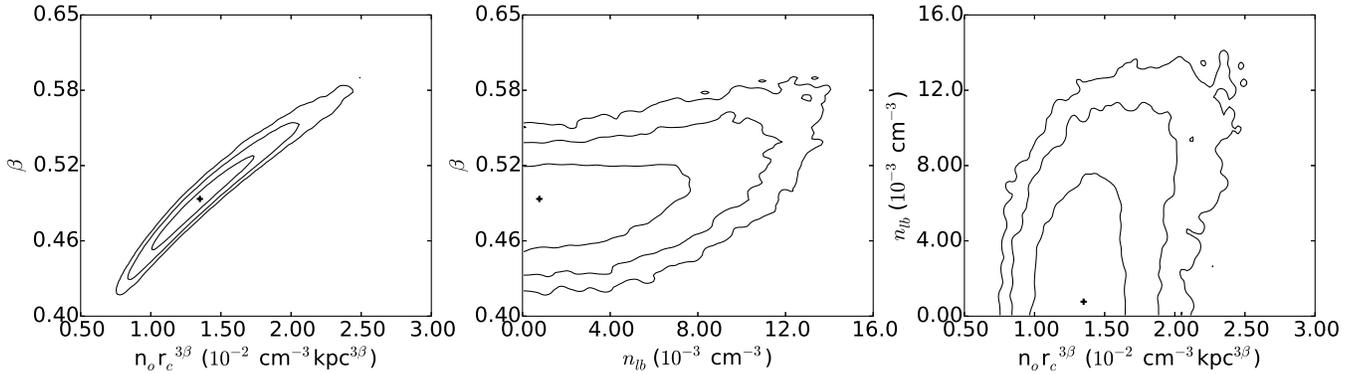}
\caption{Joint posterior probability distributions for our model parameters while fitting the \ion{O}{8} observations with an optically thin plasma.  The lines represent the 1$\sigma$, 2$\sigma$, and 3$\sigma$ confidence regions.  We trace the 1$\sigma$ boundary on the left plot (between $n_{o}r_{c}^{3\beta}$ and $\beta$) to estimate the uncertainties on halo density, mass, etc., with radius.  The black crosses represent the best-fit parameter values from the first row of Table~\ref{table.mcmc_results}.  
}
\label{figure.contours}
\end{figure}

The hot gas halo density parameters have much tighter constraints than the LB density parameter when we fit the \ion{O}{8} emission lines.  Furthermore, the LB density parameter is consistent with zero, implying that there is little \ion{O}{8} emission due to the LB.  This is seen in our marginalized posterior probability distributions (Figure ~\ref{figure.chains}) and joint probability distributions (Figure ~\ref{figure.contours}) for fitting the \ion{O}{8} lines with an optically thin hot halo plasma.  This is a somewhat self-imposed constraint based on our temperature assumption for the LB.  The \ion{O}{8} line emissivity for the LB is $\approx$50 times weaker than the halo line emissivity.  However, the LB density parameter would be much larger if there was a global contribution to the \ion{O}{8} emission lines from the LB.  The fact that the LB density parameter does not compensate for the relatively small line emissivity implies that the LB has little contribution to the \ion{O}{8}.  This means the \ion{O}{8} emission lines are effectively fit with just our hot halo model, making them a good tracer of halo gas.  

\subsection{Analyzing \ion{O}{7} Line Emission}
\label{subsection.ovii_analysis}

\begin{figure}
\centering
\includegraphics[width = 1.0\textwidth, keepaspectratio=true]{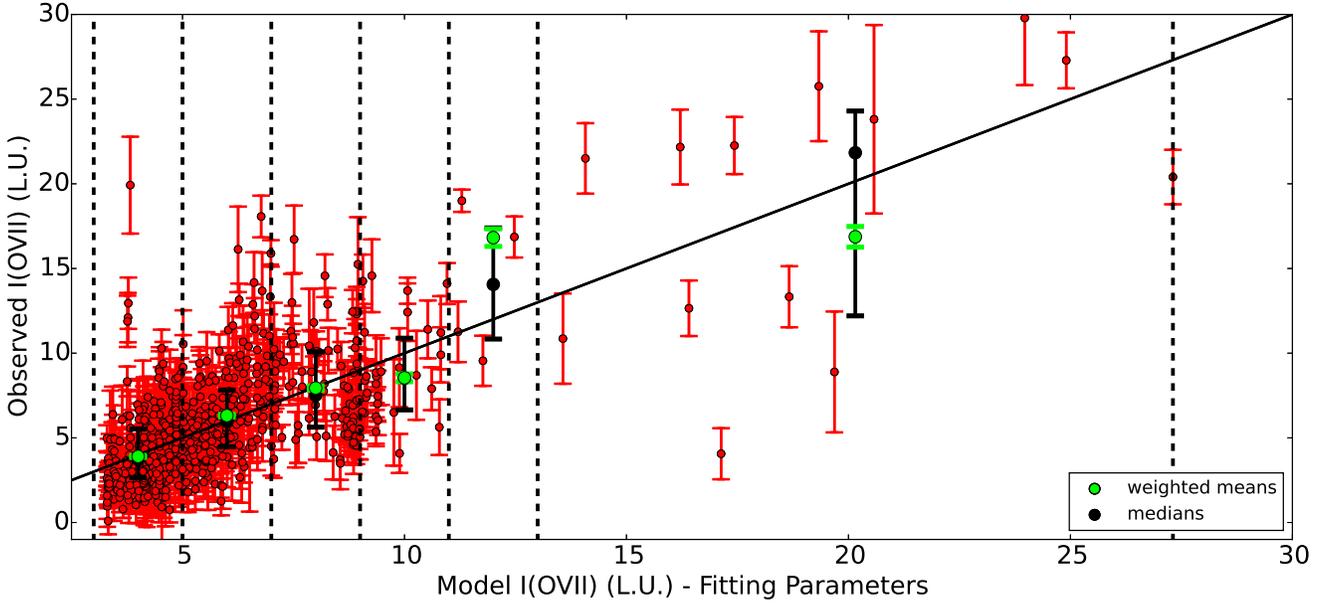}
\caption{Same as Figure~\ref{figure.obs_vs_mod} but for our \ion{O}{7} fit results.  
}
\label{figure.obs_vs_mod_o7}
\end{figure}

The \ion{O}{7} emission line observations show very different fitting results compared to the \ion{O}{8} observations.  Our fitting procedure does find optimal parameters to reproduce the data, but our best-fit model finds a $\chi_{red}^2$ (dof) = 4.69 (645) for the \ion{O}{7} observations compared to 1.08 (644) for the \ion{O}{8} observations.  Figure~\ref{figure.obs_vs_mod_o7} shows an analog of Figure~\ref{figure.obs_vs_mod}, but for our best-fit \ion{O}{7} optically thin plasma model compared to the \ion{O}{7} observations.  While the constraints on the fitted parameters are comparable to the \ion{O}{8} fitting results (even significantly better for $n_{LB}$), the quality of our fit implies that the \ion{O}{7} observations are not well described by our parametric model.

Following the approach of \cite{miller_bregman13} on \ion{O}{7} absorption lines, we fit the \ion{O}{7} emission line observations with the inclusion of an additional uncertainty to the observations ($\sigma_{add}$).  The purpose of this approach is to add the smallest uncertainty to the measurements ($\sigma_{add}$ is added in quadrature with the statistical and systematic uncertainties) while also finding an acceptable model fit to the data.  In this way, $\sigma_{add}$ is a crude estimate of the sightline-to-sightline variation or a way to quantify the deviation of the observations from our ideal density model.  We find for both an optically thin and optical depth corrected plasma, including $\sigma_{add}$ = 2.1 L.U. ($\approx$40\% of the median \ion{O}{7} line intensity and $\approx$1.7 times the median \ion{O}{7} uncertainty) results in a $\chi_{red}^2$ (dof) = 1.03 (645).  Fitting the \ion{O}{7} observations with and without this added uncertainty causes a small change in the best-fit parameters (within 1$\sigma$ of each other, see Table~\ref{table.mcmc_results}).  The source of this variation and its relation to the differences between our \ion{O}{7} and \ion{O}{8} fitting results are discussed in Section ~\ref{subsection.discrepancy}.

\subsection{Optical Depth Effects}
\label{subsection.tau_effects}

Optical depth corrections in our line intensity calculation have a subtle, yet important imprint on our model fitting results.  Regardless of whether or not we analyze \ion{O}{7} and \ion{O}{8}, applying optical depth corrections to our intensity calculation increases both our fitted halo normalization parameter and $\beta$.  This is likely due to the absorption term dominating the scattering term in Equation (\ref{eq.transfer_optical_depth}).  For a given model parameter set, the absorption term causes the calculated line intensity along a given line of sight to be smaller than the optically thin calculation (i.e., some photons are absorbed along the line of sight).  This explains the increase in the halo normalization parameter since more photons must be created everywhere in the halo to account for these absorbed photons and still reproduce the observations.  But the absorption term $\kappa \propto n$, implying that denser gas regions are more susceptible to this effect.  This explains the increase (or steepening) of the fitted $\beta$ parameter since regions closer to the Galactic center need to generate more photons compared to the outer regions of the halo to reproduce the observations.  This qualitative argument explains the behavior we see in our fit results, but we need to quantify this effect to constrain the true halo gas density profile.  

\begin{figure}
\centering
\includegraphics[width = 1.\textwidth, keepaspectratio=true]{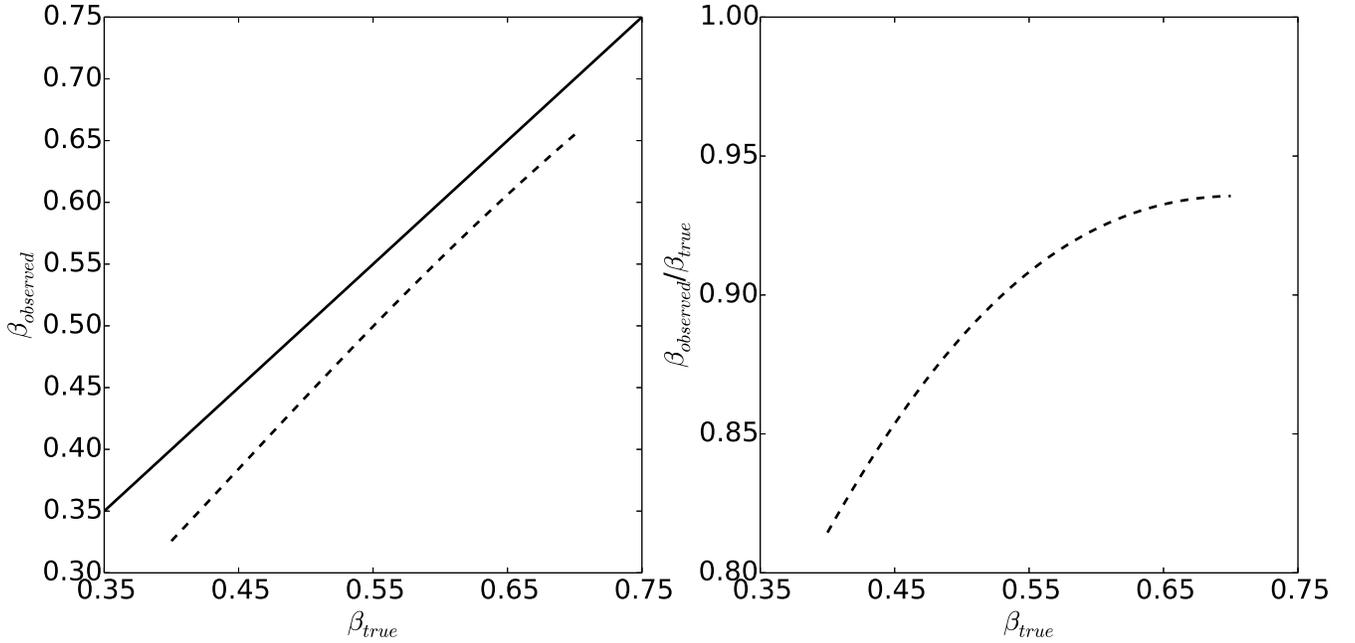}
\caption{Plots illustrating the optical depth effects we model in our fitting procedure.  We fit optical depth-corrected observations from a halo density profile with $\beta_{true}$ assuming the plasma is optically thin.  The  ``observed'' or ``fitted'' $\beta_{observed}$ values compared to the $\beta_{true}$ values are seen on the left while the right shows the ratio between the two (dashed lines).  One sees $\beta_{observed}$ ranges between 80\% and 95\% of $\beta_{true}$ for a range of true halo profiles.  
}
\label{figure.beta_map}
\end{figure}

We evaluate these optical depth corrections quantitatively by comparing the true and fitted $\beta$ parameters if the true plasma accounts for optical depth corrections, but we assume that the plasma is optically thin.  Given a true model parameter set, $(n_{o}r_{c}^{3\beta})_{true}$ and $\beta_{true}$, we simulate 160 \ion{O}{8} emission line observations randomly distributed across the sky accounting for optical depth effects.  We then fit the simulated observations assuming the plasma is optically thin.  This gives us an estimate of how much the optical depth corrections actually effect our fit parameters.  Figure~\ref{figure.beta_map} shows the results of this procedure for a true halo normalization of $1.3 \times 10^{-2}$ cm$^{-3}$ kpc$^{3\beta}$ and a range of $\beta_{true}$.  One sees if we assume the plasma is optically thin, our inferred $\beta$ would be smaller (shallower) than $\beta_{true}$.  The strength of this effect depends on $\beta_{true}$, but the observed $\beta$ ranges between 80\% and 95\% of $\beta_{true}$.  This is consistent with the differences we see in our own fit results.  

At this point it is difficult to assess the accuracy of these optical depth corrections, even with the estimation procedure above.  This is because finding solutions to the transfer equation is inherently an iterative process.  We assumed in the above calculations that our single calculation of the mean intensity for a given halo parameterization is an accurate representation of the true radiation field everywhere in the halo.  This is not necessarily the case when accounting for multiple photon scatterings, assumptions on the turbulence of the medium, etc.  Future work will involve developing a Monte Carlo radiative transfer code designed for this system so we can more accurately estimate these effects.  However, our calculation above is consistent with our expected scenario where optical depth corrections are largest toward denser regions of the halo (near the Galactic center).  Furthermore, the \ion{O}{8} best-fit parameters are consistent with each other at the 1$\sigma$ level with and without optical depth corrections.  This enhances our argument that we can indeed constrain the true density profile of the Milky Way's hot gas halo.

\section{Discussion}
\label{section.discussion}

Here we discuss the implications of our model fitting results for both \ion{O}{8} and \ion{O}{7} observations.  Our constraints on the Milky Way's hot gas halo density profile provide additional constraints on other Milky Way properties, such as its total baryonic mass, mass accretion rates, etc.  We will also discuss how our model constraints compare with previous analyses on the Milky Way's SXRB, local \ion{O}{7} absorption lines believed to be caused by the same hot gas halo plasma, and theoretical work on galactic hot halos.  

\subsection{Implications for the Milky Way}
\label{subsection.MW_implications}

\begin{figure}
\centering
\includegraphics[width = .5\textwidth, keepaspectratio=true]{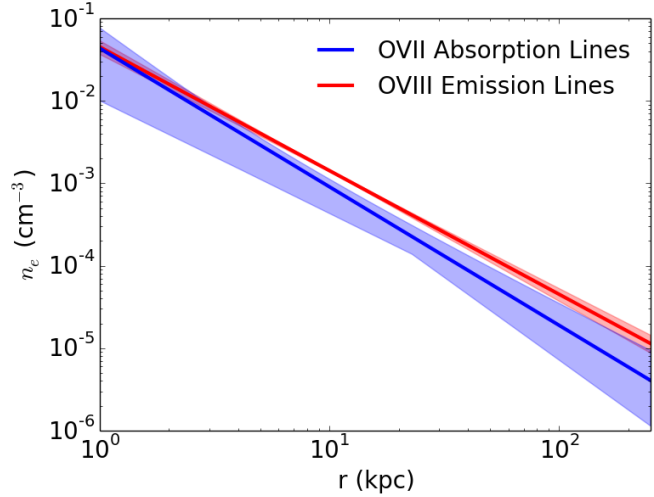}
\caption{Our best-fit density profile as a function of galactocentric radius from fitting the \ion{O}{8} observations with an optically thin plasma and assuming a gas metallicity of 0.3 $Z_{\odot}$ (red).  The blue curve shows the best-fit density profile from \cite{miller_bregman13}, who analyzed \ion{O}{7} absorption lines with a similar procedure to this work.  The shaded regions represent the 1$\sigma$ boundaries on these values.  
}
\label{figure.density}
\end{figure}

\begin{figure}
\centering
\includegraphics[width = .5\textwidth, keepaspectratio=true]{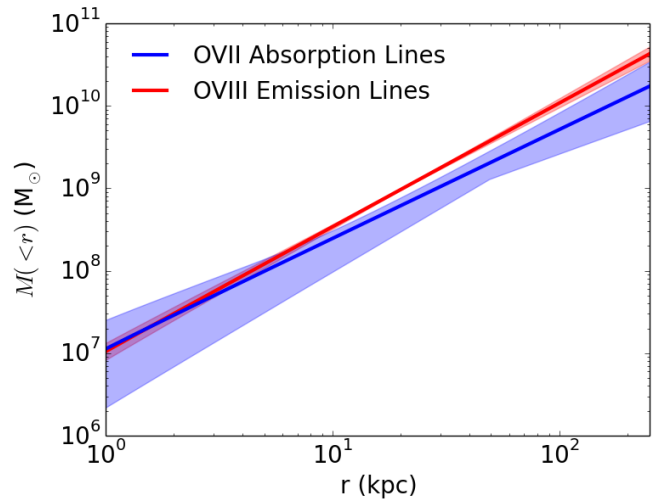}
\caption{Enclosed mass as a function of galactocentric radius for the same density profiles in Figure~\ref{figure.density}.  We find characteristic masses of the hot gas halo to be $2.9 - 5.3 \times 10^9 M_{\odot}$ within 50 kpc and $2.7 - 9.1 \times 10^{10} M_{\odot}$ within $R_{vir}$ when we examine all of our fitting procedures.  
}
\label{figure.mass}
\end{figure}

The primary quantity we estimate with our halo model constraints is the total hot gas mass within the Milky Way's virial radius ($R_{vir}$).  Estimates of the Milky Way's $R_{vir}$ range between 207 kpc \citep{loeb_etal05} and 277 kpc \citep{shattow_loeb09} with additional estimates in between \citep{klypin_etal02}.  Here we adopt a $R_{vir}$ = 250 kpc.  Figures~\ref{figure.density} and ~\ref{figure.mass} show our density and integrated mass profiles for our \ion{O}{8} optically thin plasma results assuming a gas metallicity of 0.3 $Z_{\odot}$ (red curves).  The shaded regions indicate the 1$\sigma$ mass boundaries by tracing the 1$\sigma$ contour in $n_{o}r_{c}^{3\beta} - \beta$ space (see Figure~\ref{figure.contours}).  Table~\ref{table.mcmc_results} also shows the integrated hot gas mass out to 50 and 250 kpc for every observation/plasma combination we examine.  Our model fitting results imply hot gas masses between $2.9 - 5.3 \times 10^9 M_{\odot}$ within 50 kpc and $2.7 - 9.1 \times 10^{10} M_{\odot}$ within $R_{vir}$.  These estimates are more tightly constrained if we only consider our \ion{O}{8} fitting results, which reproduce the data much better than our \ion{O}{7} fitting results.  The characteristic masses have a range of $2.9 - 3.8 \times 10^9 M_{\odot}$ within 50 kpc and $2.7 - 4.7 \times 10^{10} M_{\odot}$ within 250 kpc in this case.

We compare these hot gas mass estimates to the known baryonic and total mass of the Milky Way.  The known baryonic mass in the Milky Way (stars + cold gas + dust) is between $6-7 \times 10^{10} M_{\odot}$ \citep{binney_tremaine08}.  If our hot gas density profile extends out to $R_{vir}$, the hot gas and known baryonic masses are comparable.  We can extend these mass constraints to estimate the Milky Way's total baryon fraction, defined here as $f_b \equiv M_b / M_{tot}$, where $M_b$ is the total baryon mass and $M_{tot}$ is the total baryon plus DM mass .  Current estimates for the Milky Way's virial mass have a range of $1.0 - 2.4 \times 10^{12} M_{\odot}$ \citep{boylan_kolchin_etal13}.  If we account for this range of virial masses, our range of hot gas masses from our \ion{O}{8} observations (with 1$\sigma$ uncertainties), we estimate the Milky Way's total baryon fraction to be between 0.03 and 0.11.  Even the upper limit of this range (which makes numerous assumptions on the various mass estimates involved) falls well below the cosmological baryon fraction of $f_b = 0.171 \pm 0.006$ measured by the \textit{Wilkinson Microwave Anisotropy Probe} \citep{dunkley_etal09}.  This implies that if our best-fit hot halo density profile extends to the $R_{vir}$, it contains a significant amount of mass compared to the other baryons in the Milky Way but does not account for all of the Milky Way's ``missing baryons''.  

It is possible our hot gas density profile extends past $R_{vir}$, thus increasing our mass estimates.  In particular, we estimate how far our density profile would need to extend to account for all the Milky Way's missing baryons.  The Milky Way's virial mass, the other known baryonic mass components (excluding the hot gas mass), and the cosmological baryon fraction discussed above imply a missing baryon mass of $\sim 2 \times 10^{11} M_{\odot}$.  Given the range of assumptions made above for our hot gas mass estimates above (gas metallicity, mass uncertainties, etc.), our halo would need to extend between 3 and 5 $R_{vir}$ to account for all of the Milky Way's missing baryons \citep[][in preparation]{bregman_etal15}.  

The hot mass estimates quoted in Table~\ref{table.mcmc_results} all assume a gas metallicity of 0.3 $Z_{\odot}$ based on pulsar dispersion measurements toward the LMC.  This is a powerful constraint in our analysis since it probes the gas properties out to the fixed distance of the LMC, $\approx$50 kpc.  Here we define the dispersion measure as

    \begin{equation}
     DM = \int_{0}^{d} n_e(s) ds,
    \end{equation}

\noindent where $d$ is the distance to the source and $n_e(s)$ is the density profile along the line of sight.  \cite{anderson_bregman10} examined numerous DM measurements for sources associated with the LMC.  After accounting for DM contributions from the Milky Way's disk and from the LMC itself, they estimated a DM for the Milky Way's hot halo of 23 cm$^{-3}$ pc (\cite{fang_etal13} conducted a similar analysis and found similar results).  Integrating our \ion{O}{8} optically thin halo gas model in the LMC direction ($l,b$ = 273.57$\arcdeg$, -32.08$\arcdeg$) yields a DM of 8.2 cm$^{-3}$ pc.  However, the calculated DM $\propto$ $Z^{-1}$ and the line emissivity values used in our analysis assumed \cite{anders_grevesse89} solar abundances, or equivalently a solar metallicity.  This means that we match the hot halo DM estimate from \cite{anderson_bregman10} if we assume a halo gas metallicity of $Z$ = 0.3 $Z_{\odot}$.  This estimate assumes that our hot gas halo density profile accounts for all of the residual DM, which may not be the case if there are other unknown electron sources along our line of sight.  We thus present this estimate as a lower limit to the hot halo characteristic gas metallicity, $Z$ $\geq$ 0.3 $Z_{\odot}$.  

\begin{figure}
\centering
\includegraphics[width = .5\textwidth, keepaspectratio=true]{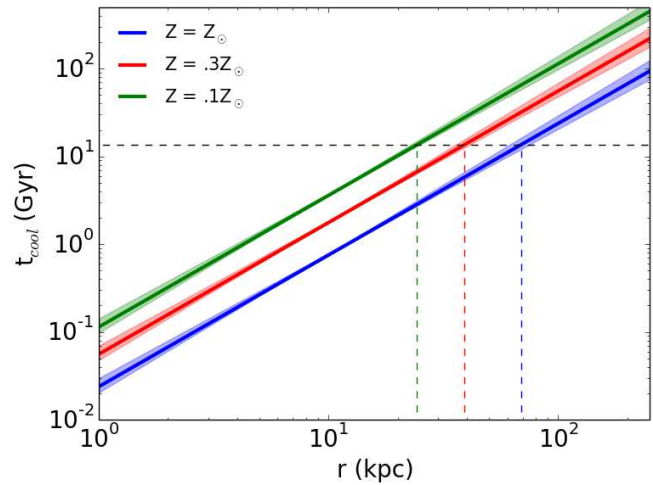}
\caption{Cooling time as a function of radius calculated using Equation (\ref{eq.tcool}) for the density profile in Figure~\ref{figure.density}.  The different colors represent different gas metallicities with more metals resulting in shorter cooling times.  The black horizontal line represents the age of the universe (13.6 Gyr) and the colored dashed lines represent the cooling radii for different metallicities (between 25 and 70 kpc).
}
\label{figure.tcool}
\end{figure}

We also examine the thermal properties of the halo gas given our range of best-fit models.  Quantities such as the cooling time and corresponding cooling radius for the halo gas offer insight on whether or not our hot gas model is stable at this moment in the Milky Way's evolution.  We adopt the expression for the cooling time from \cite{fukugita_peebles06}:

	\begin{equation}
	 \label{eq.tcool}
	 t_{cool}(r) = \frac{1.5nkT}{\Lambda(T,Z) n_e(n - n_e)} \approx \frac{1.5kT\times 1.92}{\Lambda(T,Z) n_e\times 0.92},
	\end{equation}

\noindent where $\Lambda (T,Z)$ is the bolometric cooling rate as a function of temperature and metallicity \citep{sutherland_dopita93}.  Figure~\ref{figure.tcool} shows the cooling time as a function of radius for our best-fit \ion{O}{8} optically thin plasma model.  The colors represent cooling times for different metallicities and the shaded boundaries are calculated in the same way as the mass boundaries in Figure~\ref{figure.mass}.  

One useful quantity we estimate from these cooling times is the hot gas halo cooling radius, or where $t_{cool}$ = 13.6 Gyr.  The cooling radius is a potential constraint on the inner radius to our halo profile since for $t_{cool} \textless$ 13.6 Gyr for $r \textless r_{cool}$, implying that the halo gas would be in a cooler phase at this point in the Milky Way's evolution in the absence of additional heating sources.  The dashed lines in Figure~\ref{figure.tcool} show the instantaneous $r_{cool}$ for metallicities ranging between 0.1 and 1.0 $Z_{\odot}$, assuming no extra thermal energy has been added to the gas.  The range of metallicities result in $r_{cool}$ between 25 and 70 kpc, with 40 kpc being the cooling radius if $Z$ = 0.3 $Z_{\odot}$.  We emphasize these cooling radius estimates should not be taken as literal estimates for an inner gas halo radius since they completely ignore energy input that is likely present in the form of SNe \citep{maclow_etal89, deavillez_maclow01} or possibly AGNs \citep{su_etal10} in the Milky Way.  These estimates do characterize the physical state of the gas as it currently exists and are important for estimating other quantities related to the cooling time of the gas.  

\begin{figure}
\centering
\includegraphics[width = .5\textwidth, keepaspectratio=true]{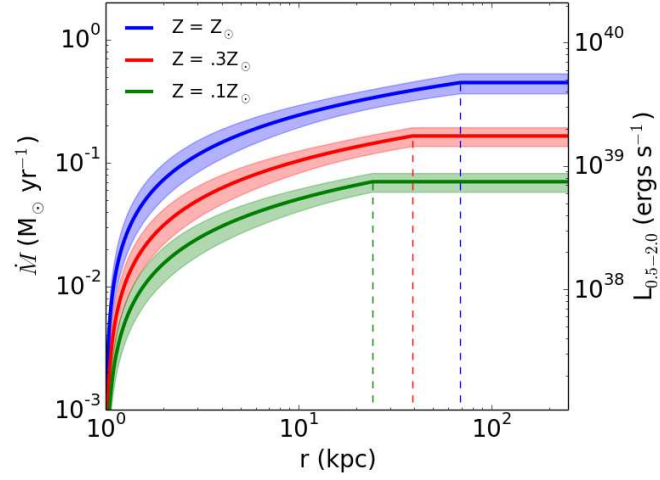}
\caption{Integrated mass accretion rate calculated using Equation (\ref{eq.mdot}) for the density profile and cooling times in Figures~\ref{figure.density} and ~\ref{figure.tcool}.  The colors and dashed lines are also the same as in Figure~\ref{figure.tcool}.  We find mass accretion rates $\lesssim$ 0.5 $M_{\odot}$ yr$^{-1}$, less than the Milky Way's SFR.  The right axis of the plot also shows the conversion between $\dot{M}$ and $L_X$ in the 0.5 - 2.0 keV band (Equation (\ref{eq.luminosity})).  
}
\label{figure.mdot}
\end{figure}

The hot gas cooling time is directly related to the cooling rate and corresponding hot gas halo mass accretion rate.  These estimates offer constraints on the halo gas as it cools into a cooler phase gas and eventually falls back on to the Milky Way's disk to eventually form stars.  We calculate the current integrated mass accretion rate by integrating the mass within spherical shells divided by the cooling time at the shells' radii.  This is represented by the following equation:

    \begin{equation}
     \label{eq.mdot}
     \dot{M}(r) = \int_{0}^{r} \frac{\mu m_{p} n_{halo}(r')}{t_{cool}(r')} 4 \pi r'^{2} dr',
    \end{equation}

\noindent where $\mu$ = 0.59 is the mean mass per particle and $m_p$ is the proton.  Figure~\ref{figure.mdot} shows our integrated mass accretion rate profiles for the same density and cooling time profiles discussed above.  Like Figure~\ref{figure.tcool}, the dashed lines here also represent the cooling radii for each metallicity curve we show.  Note the curves flatten for $r \textgreater r_{cool}$ because we calculate the \textit{current} mass accretion rate.  In this case, regions where $t_{cool} \textgreater$ 13.6 Gyr have not had time to cool at this point.  These results imply that the hot gas halo loses between 0.08 and 0.50 $M_{\odot}$ yr$^{-1}$ in the inner regions of the halo as the gas cools.  The upper limits of these accretion rates are similar to simulated accretion rates of halo gas around Milky Way analogs \citep{fraternali_binney08}.  The upper limits here also imply accretion of halo gas may be an important contributor to the Milky Way's current star formation rate (SFR), observed to be between 0.68 and 1.45 $M_{\odot}$ yr$^{-1}$ \citep{robitaille_whitney10}.  Alternatively, \cite{leitner_kravtsov11} modeled the mass-loss rates and star formation histories for a sample of galaxies (including the Milky Way) to examine if recycled gas from stellar mass loss could sustain observed levels of star formation in the galaxies.  They concluded that mass loss from later stages of stellar evolution could provide most of the fuel required to produce the observed Milky Way SFR.  This favors a sub-solar halo gas metallicity such that the combination of stellar mass loss and our estimates for accretion from the CGM do not overproduce the observed Milky Way's SFR.

The final quantity we calculate related to the thermal properties of the halo gas is the halo X-ray luminosity ($L_X$).  Specifically, we estimate the 0.5 - 2.0 keV band luminosity to compare with results from the RASS \citep{snowden_etal97}.  The 0.5 - 2.0 keV band luminosity is related to the cooling rate (or mass accretion rate) by the following conversion:

    \begin{equation}
     \label{eq.luminosity}
     L_X (r) = 0.412 \times \dot{M}(r) \frac{1.5 kT}{\mu m_p},
    \end{equation}

\noindent where $\dot{M}(r)$ is the accretion rate from Equation (\ref{eq.mdot}) and the constant 0.412 is the conversion between bolometric to 0.5 - 2.0 keV flux using WEBPIMMS \footnote{\url{http://heasarc.gsfc.nasa.gov/cgi-bin/Tools/w3pimms/w3pimms.pl}}.  Figure~\ref{figure.mdot} shows this conversion, represented by the scaling on the right side of the figure.  Our calculated 0.5 - 2.0 keV band luminosity has a range of $0.8 - 5.0 \times 10^{39}$ erg s$^{-1}$.  Previous works have modeled the \textit{ROSAT} 3/4 keV background with non-spherical geometrical models and arrived at similar luminosities.  \cite{snowden_etal97} modeled the X-ray bulge emission as a cylinder with a radius of 5.6 kpc and density scale height of 1.9 kpc to find a bulge 0.5 - 2.0 keV luminosity of $\sim 2 \times 10^{39}$ erg s$^{-1}$.  Similarly, \cite{wang98} modeled the all-sky \textit{ROSAT} emission with a axisymmetric, disk-like geometry and found a 0.5 - 2.0 keV luminosity of $\sim 3 \times 10^{39}$ erg s$^{-1}$.  Our calculated 0.5 - 2.0 keV luminosity is consistent with these estimates of $2-3 \times 10^{39}$ erg s$^{-1}$ for an assumed metallicity of 0.3 $Z_{\odot}$.  This is another indication that the halo gas metallicity is sub-solar, with $Z$ = 0.3 $Z_{\odot}$ being a limit that satisfies numerous observational constraints.

\subsection{Comparing with Previous Observational Work}
\label{subsection.previous_work_obs}

Our primary comparison with previous work on the Milky Way's hot gas halo follows the work by \cite{miller_bregman13} on \ion{O}{7} absorption lines in QSO spectra.  Other works on different observations with different analyses will be addressed, but we focus on this work since the procedure for fitting and analyzing the absorption lines in \cite{miller_bregman13} is identical to our work.  Both of these works model the observations with a hot gas halo model defined by a modified $\beta$-model (Equation (\ref{eq.beta_model_approx})).  The obvious difference between these works is the type of observation analyzed, \ion{O}{7} column densities/equivalent widths ($\propto nL$) in \cite{miller_bregman13} compared to \ion{O}{7}/\ion{O}{8} emission line intensities ($\propto n^{2}L$) here.  The physical differences between the absorption and emission line density scalings imply that the similarities or differences between the results tell us about the structure of the hot gas halo.

The model fitting results from \cite{miller_bregman13} yielded hot gas halo structural parameters of $n_{o}r_{c}^{3\beta}$ = $1.3^{+1.6}_{-1.0} \times 10^{-2}$ cm$^{-3}$ kpc$^{3\beta}$, $\beta$ = $0.56^{+0.10}_{-0.12}$, $\chi^{2}$ (dof) = 31.0 (26) assuming an optically thin plasma and $n_{o}r_{c}^{3\beta}$ = $4.8^{+8.5}_{-3.7} \times 10^{-2}$ cm$^{-3}$ kpc$^{3\beta}$, $\beta$ = $0.71^{+0.17}_{-0.20}$, $\chi^{2}$ (dof) = 26.0 (27) accounting for saturation of the absorption lines.  Both of these results include added uncertainties to the equivalent width observations of 7.5 and 7.2 m\AA\ for the optically thin and optical depth correction cases to find acceptable $\chi^{2}$ values.  There are several important comparisons to make between these absorption line results and our fit results from the emission lines in Table~\ref{table.mcmc_results}.  

We point out that the uncertainties on the absorption line fit parameters are much larger than any of our emission line fitting results.  This is likely due to the sample size for the emission lines being $\approx$20 times larger than the absorption line sample.  Even with the relatively small uncertainties on the emission line fit results, the fit parameters  are consistent with each other based on their 1$\sigma$ uncertainties.  This consistency applies when comparing the optically thin results separately from the optical depth corrected results.  The only exception is the \ion{O}{7} emission line result with optical depth corrections, where we report a shallower $\beta$ (0.47$\pm$0.01) than the absorption line results.  This discrepancy is likely due to a combination of our treatment of optical depth corrections (for both the absorption and emission lines) and the overall variation we see in the \ion{O}{7} emission lines.  The fact that the model fitting results are consistent with each other allows us to compare additional quantities based on our halo models.  

We provide an additional estimate of the halo gas metallicity (or oxygen abundance) independent from the LMC pulsar dispersion measure constraint.  This estimate utilizes the ratio between the absorption and emission line density profile results.  The \ion{O}{7} absorption lines/column densities probe $n_{OVII}(r) = X_{OVII}n_{ox}(r)$, where $X_{OVII}$ = 0.5 is the \ion{O}{7} ion fraction and $n_{ox}(r)$ is the oxygen density profile.  On the other hand, the emission lines are sensitive to $n_e(r)^2$ or $n_H(r)^2$ since $I \propto \int n_e n_{ion} ds \propto \int n_e^2 X_{ion} Z ds$ (see Equation (\ref{eq.intensity_thin})).  Thus, the ratio between the absorption and emission line density profiles is a direct estimate of the gas metallicity distribution, $n_{ox}(r) / n_H(r)$.  Here, we take the optically thin electron density distribution from \cite{miller_bregman13} and convert it to an oxygen density distribution using a solar oxygen abundance of log$(N_O)$ = 8.74 from \cite{holweger_01}.  For the hydrogen density profile, we use our \ion{O}{8} optically thin model fitting results in Table~\ref{table.mcmc_results} and assume $n_H = 0.8 n_e$ (note the results in Table~\ref{table.mcmc_results} are for electron densities).  When we divide these two density profiles, we find a weak halo gas metallicity gradient of approximately $Z \propto r^{-0.2}$ with a metallicity of $Z = 0.26 \pm 0.10$ $Z_{\odot}$ (1$\sigma$) at 10 kpc for \cite{holweger_01} solar abundances.  These results are also consistent with the pulsar dispersion measure gas metallicity constraint of $Z$ $\geq$ 0.3 $Z_{\odot}$ discussed in Section ~\ref{subsection.MW_implications}.

The most important quantity we compare between this work and the \cite{miller_bregman13} results is the hot gas mass profile.  In addition to the best-fit model results in this paper, Figures~\ref{figure.density} and ~\ref{figure.mass} show the density and mass curves from \cite{miller_bregman13} assuming an optically thin plasma (red shaded area).  Like the best-fit parameters, the total mass estimates within $R_{vir}$ are consistent with each other based on the 1$\sigma$ uncertainties.  This implies characteristic hot gas halo masses of $\approx 2-5 \times 10^{10}$ $M_{\odot}$ within $R_{vir}$ regardless if one analyzes the emission or absorption lines.  

We compare these mass estimates and underlying density profiles with other independent absorption line studies on the Milky Way's hot gas halo.  The comparisons here are limited since previous work on the X-ray absorption lines has either been confined to individual sightlines \citep[e.g.,][]{rasmussen_etal03} and/or compared an ensemble of absorption lines with individual or small samples of emission line measurements \citep{bregman_ld07, gupta_etal12}.  The latter implies that results on the aggregate absorber properties may not be consistent with the aggregate emission properties while the former results are only valid for individual lines of sight.  

Examples of detailed analyses of local X-ray \ion{O}{7} and \ion{O}{8} absorption include \cite{yao_etal09_b}, who analyzed the LMC X-3 sightline, and \cite{hagihara_etal10}, who analyzed the PKS 2155-304 sightline.  Both studies assume the absorbers are due to a local Galactic disk density model and fit their observations with scale heights of $\approx$2 kpc.  These result in mass estimates of $\sim 10^{8} M_{\odot}$, significantly lower than our expected values.  However, these exponential disk models tend to overproduce the Milky Way's X-ray surface brightness profile \citep{fang_etal13}, whereas our results are consistent with this constraint.  

Alternatively, the analyses by \cite{bregman_ld07} and \cite{gupta_etal12} assume that the absorption lines come from a more extended hot gas halo medium.  \cite{bregman_ld07} compared the aggregate column densities of 26 high S/N local \ion{O}{7} absorbers with a single emission measure estimate for the Milky Way's hot halo \citep{mccammon_etal02}.  \cite{gupta_etal12} took a similar approach, but they used a smaller sample of absorption lines (8 targets) and an average hot halo emission measure from multiple SXRB observations distributed across the sky (14 measurements from \cite{yoshino_etal09} and 26 measurements from \cite{henley_etal10}).  Both works characterize the Milky Way's hot gas halo as a constant density sphere of size $L$ by comparing an average column density ($\propto nL$) with an average emission measure ($\propto n^{2}L$).  The two studies disagree, though, where \cite{bregman_ld07} find $n_e = 9 \times 10^{-4}$ cm$^{-3}$, $L$ = 20 kpc, $M(\textless L) = 4 \times 10^{8} M_{\odot}$ and \cite{gupta_etal12} find $n_e = 2.0 \times 10^{-4}$ cm$^{-3}$, $L \textgreater$ 139 kpc, $M \textgreater 6.1 \times 10^{10} M_{\odot}$.  The differences come from a number of effects, including the measurement procedure of the equivalent widths and the conversion between equivalent widths and column densities.  Regardless of these differences, the constant density sphere models used in these works are not physically motivated.  One expects the gas density to decrease with distance away from the Milky Way's center if it is approximately in hydrostatic equilibrium with the Milky Way's DM distribution.  We emphasize the benefit of our procedure since we analyze the X-ray emission and absorption lines assuming the same type of density distribution that decreases with galactocentric radius.

Similar to absorption line studies on the Milky Way's hot gas halo, we also compare our results to studies focusing exclusively on the hot gas halo emission properties.  These works also range from detailed analyses of the SXRB on individual sightlines \citep[e.g.,][]{mccammon_etal02, smith_etal07} to analyses on the global emission properties of the SXRB \citep[e.g.,][]{yoshino_etal09, hs13}.  Our work offers an improvement over these previous works since we are using the largest data set to date to characterize the line emission properties of the hot gas halo with a physical model.  We nonetheless compare our results as a consistency check with previous work.  

As simple consistency checks, we compare our best-fit model line intensities and emission measures to measured values along individual sightlines.  \cite{mccammon_etal02} analyzed a $\sim$1 sr region of the sky at $l,b$ = 90$\arcdeg$, 60$\arcdeg$ using a quantum calorimeter sounding rocket.  The sensitivity and spectral resolution of their detectors allowed them to produce precise measurements of the SXRB, including an absorbed component emission measure (EM) of $3.7 \times 10^{-3}$ cm$^{-6}$ pc.  We compute our best-fit \ion{O}{8} optically thin model EM in this region and find a range of values covering the 1 sr field of view of 1.6 - 3.8 $\times 10^{-3}$ cm$^{-6}$ pc with a solid angle-weighted value of $2.6 \pm 0.2 \times 10^{-3}$ cm$^{-6}$ pc.  This implies that our computed model EM broadly comports with the observed value and does not overproduce the non-local emission in this region of the sky.  While this is only one individual line of sight observation against which we compare our model, our work is designed to characterize the global properties of the hot halo component of the SXRB.  

There have been limited studies on the emission properties of the Milky Way's hot gas halo using a large sample of sightlines.  These works typically follow a similar fitting procedure to that outlined in Section ~\ref{subsection.line_measurements}, except one fits the spectrum with two APEC models for the LB and hot gas halo emission components.  \cite{yoshino_etal09} analyzed SXRB spectra in nine fields of view with \textit{Suzaku}, but limited their sky coverage to 65$\arcdeg \textless l \textless 295 \arcdeg$.  Although this sample presents limited sky coverage, they report a range of hot halo EM values between $\approx .5-5 \times 10^{-3}$ cm$^{-6}$ pc.  \cite{hs13} presented a similar analysis on 110 \textit{XMM-Newton} observations, but also limited their observation selection to $|b| \textgreater 30 \arcdeg$ (among other selection criteria).  They report a similar range in EM from $\sim 0.4-7 \times 10^{-3}$ cm$^{-6}$ pc with a median detection of $1.9 \times 10^{-3}$ cm$^{-6}$ pc.  We calculate our best-fit \ion{O}{8} optically thin EM for the same set of sightlines as \cite{hs13} and find a range of $\sim 1-7 \times 10^{-3}$ cm$^{-6}$ pc with a median of $2.2 \times 10^{-3}$ cm$^{-6}$ pc.  These model values are also consistent with the work from both \cite{yoshino_etal09} and \cite{hs13}.  

We have shown in the above discussion our model fitting results are generally consistent with previous observational results on the local X-ray absorption and emission line observations.  The methods in previous works range from detailed studies on individual sightlines to analyzing the global properties of either the emission or absorption lines in question.  The physical interpretation of the emission and absorption lines also ranges from a compact exponential disk of hot gas material to more extended distributions.  We have attempted to unify this picture by analyzing both a large sample of emission and absorption line observations independently, but with the same model fitting procedure.  With these comparisons, we have shown that our modified $\beta$-model for an extended hot gas halo density profile can describe both the emission and absorption line observations, but is also broadly consistent with previous independent projects on the various types of observations.

\subsection{Comparing with Simulations}
\label{subsection.previous_work_theory}

We compare our hot gas density model results with simulations of isolated Milky-Way-sized halos.  Specifically, we focus on simulations designed to analyze hot diffuse halo gas in the galaxies.  There are many details to compare between our parametric model results and these simulations, but here we compare the most basic properties of the halo gas: the density estimates at different radii, the mass contained in hot gas, and the gas metallicity.  

\begin{figure}
\centering
\includegraphics[width = .5\textwidth, keepaspectratio=true]{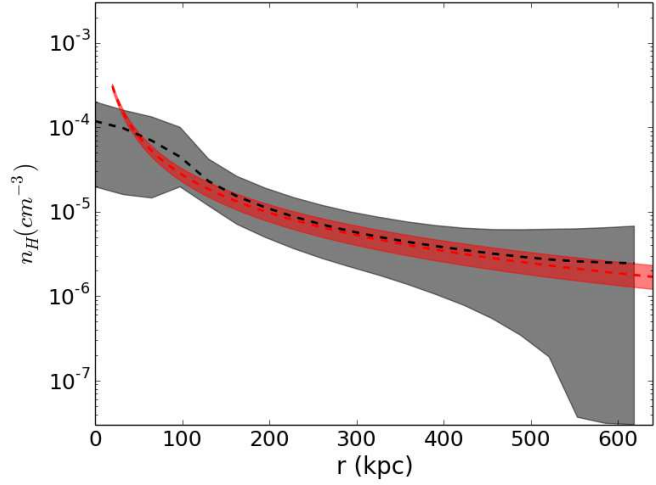}
\caption{Our best-fit model density profile (red) compared to a recent suite of simulations from \cite{nuza_etal14}.  They calculate the hot gas density profile for many random projections with the black dashed line and shaded region representing the mean and standard deviation of their calculations.  We find excellent agreement with these simulations for $r \gtrsim 50$ kpc.  
}
\label{figure.nuza}
\end{figure}

Our best-fit halo density model is consistent with recent simulations on the Local Group system.  \cite{nuza_etal14} analyzed the distribution of all gas phases in a suite of simulations of the Local Group as part of the Constrained Local Universe Simulations project.  Figure~\ref{figure.nuza} shows our best-fit halo density profile compared to their hot gas ($T \geq 10^{5}$ K) density profile for the Milky Way analog.  Even though their simulations show some variation (the gray shaded region represents density profile estimates from random viewing angles of their simulated galaxy), the two density profiles are remarkably similar for $r \gtrsim 50$ kpc.  The discrepancy within $\approx$50 kpc is likely due to feedback mechanisms within the galactic disk and contributions from gas $\textless 10^6$ K.  This recent result indicates that the structure of our model hot gas halo is qualitatively similar to current simulations.

We compare our best-fit halo mass estimates within $R_{vir}$ to hot gas halo simulations.  These simulations are all designed to analyze different aspects of their host galaxies, but typically estimate hot gas masses with a range of 10$^{10}$ - 10$^{11}$ $M_{\odot}$.  For example, the \cite{nuza_etal14} simulations discussed above were designed to analyze gas properties in the Local Group medium and report hot gas masses between $4$ and $5 \times 10^{10} M_{\odot}$ within $R_{vir}$ of their Milky Way analog.  Another example by \cite{marinacci_etal14} utilized the moving-mesh code \texttt{AREPO} to analyze the relationship between stellar feedback processes and diffuse gas in galaxies.  They report a wider range of warm plus hot gas masses between 10$^{10}$ - 10$^{11}$ $M_{\odot}$.  $N$-body simulations from \cite{sommer_larsen06} focused specifically on the hot gas mass contained within $R_{vir}$ of Milky-Way-type galaxies and found hot gas masses comparable to the stellar masses of the host galaxies ($\approx 2.3 \times 10^{10} M_{\odot}$).  This result is qualitatively similar with our hot gas mass estimates compared to the Milky Way's stellar mass.  In an entirely different approach, \cite{feldmann_etal13} used hydrodynamical simulations to analyze cosmic rays scattering off the Milky Way's hot gas halo and their observational signatures in the diffuse gamma-ray background \citep{keshet_etal04}.  They report \ion{H}{2} gas masses of [0.2, 1.0, 3.5] $\times 10^{10} M_{\odot}$ for $r$ = [50, 100, 200] kpc.  These results indicate that our best-fit model hot gas mass estimates between $2.7$ and $4.7 \times 10^{10} M_{\odot}$ within 250 kpc are consistent with current simulations of galaxy formation.  

Our lower limit on the halo gas metallicity of $\geq$0.3 $Z_{\odot}$ is consistent with simulations as well.  The halo gas metallicity tends to have more variation in the literature depending on the simulation and the investigated galaxy evolution properties (typically feedback mechanisms).  For example, hydrodynamical simulations analyzing hot gas halos as shock-heated material accreted on the DM potential wells of their galaxies favor sub-solar halo gas metallicities of $\lesssim$0.5 $Z_{\odot}$ \citep{toft_etal02, cen_ostriker06, cen12}.  On the other hand, the \cite{marinacci_etal14} simulations discussed above were designed to analyze stellar feedback properties on the hot gas.  These results predict a metallicity gradient in the halo gas starting $\sim$1 $Z_{\odot}$ near the galaxy disk and dropping below $\sim$0.3 $Z_{\odot}$ for $r \gtrsim 20$ kpc.  The lower limit we place on the Milky Way's halo gas metallicity of $\geq$0.3 $Z_{\odot}$ implies a minimal level of enrichment from the Milky Way's disk.

\subsection{\ion{O}{8} - \ion{O}{7} Discrepancy}
\label{subsection.discrepancy}

The \ion{O}{7} and \ion{O}{8} observations yield 2-3$\sigma$ discrepancies for our best-fit model parameters (see Table~\ref{table.mcmc_results}).  For example, the difference between our \ion{O}{7} and \ion{O}{8} fitted $\beta$ with optical depth corrections and with an added uncertainty to the \ion{O}{7} observations is 0.09, a 2.6$\sigma$ discrepancy.  Not only are these differences statistically significant, but the \ion{O}{7} fitting results yield systematically smaller halo gas normalization and $\beta$ parameters compared to the \ion{O}{8} fitting results.  Moreover, our best-fit models for the \ion{O}{7} observations consistently yield unacceptable $\chi_{red}^2$ values ($\chi_{red}^{2}$ (dof) = 4.7 (645)), whereas the \ion{O}{8} observations are well described by our parametric model ($\chi_{red}^{2}$ (dof) = 1.1 (644)).  Although these discrepancies do not significantly affect our mass estimates, they contain additional information on the physical properties of the hot halo or LB.

The difference in the fitting parameters may be evidence for a radial temperature gradient in the Milky Way's hot gas halo.  The \ion{O}{7} and \ion{O}{8} volumetric line emissivities peak at similar temperatures of log($T$) = 6.3 and 6.5 respectively \citep{sutherland_dopita93, foster_etal12}, implying the halo gas \ion{O}{7} and \ion{O}{8} emission lines arise from a single plasma with the same density and temperature profile.  To address this constraint, we note our model line intensities depend both on density and temperature ($dI \propto n^2 \epsilon(T)$ for every location along each line of sight).  Thus, changes to our flat temperature distribution would change our best-fit density profiles to produce the same model line emission.  We represent these changes by the following equation:

    \begin{equation}
     \label{eq.dI_conserved}
     n_{ion, flat}^2(r) \epsilon_{ion}(T_{flat})
     =
     n_{ion, grad}^2(r) \epsilon_{ion}(T_{grad}(r)),
    \end{equation}
    
\noindent where the \textit{flat} subscripts refer to our isothermal halo model fitting results and the \textit{grad} subscripts refer to the corrected density profile with a new temperature gradient.  

\begin{figure}
\centering
\includegraphics[width = 1.\textwidth, keepaspectratio=true]{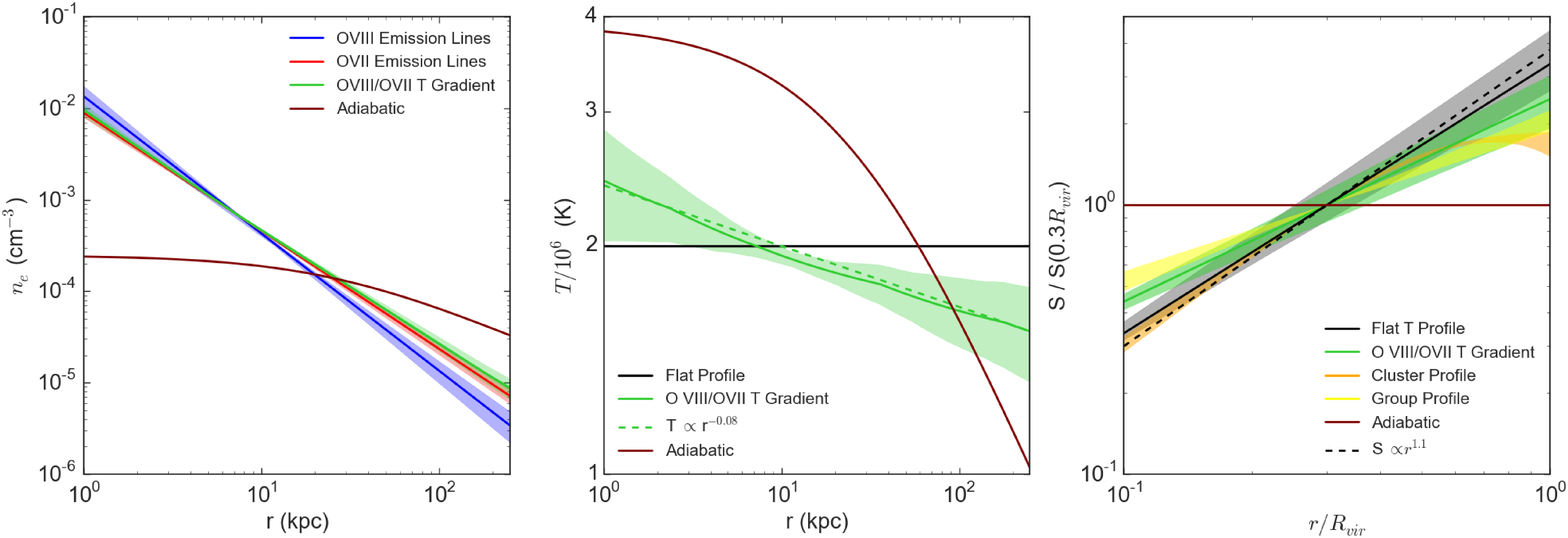}
\caption{
Left: density profiles from fitting the \ion{O}{7} (red) and \ion{O}{8} (blue) emission observations with an optically thin plasma.  The green line is the corrected density profile using the inferred temperature gradient from the red and blue density profiles.  The maroon line is the adiabatic halo density profile used by \cite{maller_bullock04} and \cite{fang_etal13}.  
Center: initial isothermal temperature profile (black line) compared to the corrected halo temperature profile using the \ion{O}{7} and \ion{O}{8} model fitting results (green solid line).  The green dashed line represents a $T \propto r^{-0.08}$ slope for reference.  The maroon line again represents the adiabatic halo gas temperature profile.  
Right: entropy profiles scaled by $S(0.3R_{vir})$ for our initial flat temperature fitting results (black solid line), corrected temperature and density model (green line), galaxy groups (yellow shaded region), and galaxy clusters (orange shaded region).  The black dashed line represents an $S \propto r^{1.1}$ slope (see text for details) while the maroon line is a flat, or adiabatic entropy profile.  
}
\label{figure.temp_grad}
\end{figure}

We estimate temperature and corrected density profiles of the halo gas using the ratio between our best-fit optically thin models for the \ion{O}{8} and \ion{O}{7} observations separately.  We utilize Equation (\ref{eq.dI_conserved}) for our \ion{O}{8} and \ion{O}{7} model fitting results to estimate a temperature gradient as

    \begin{gather}
    \begin{split}
     \label{eq.t_grad}
     \frac{n_{OVIII, flat}^2(r) \epsilon_{OVIII}(T_{flat})}
          {n_{OVII,  flat}^2(r) \epsilon_{OVII} (T_{flat})}
     &= \\
     \frac{n_{OVIII, grad}^2(r) \epsilon_{OVIII}(T_{grad}(r))}
          {n_{OVII,  grad}^2(r) \epsilon_{OVII} (T_{grad}(r))}
     &=
     \frac{\epsilon_{OVIII}(T_{grad}(r))}
          {\epsilon_{OVII} (T_{grad}(r))},
    \end{split}
    \end{gather}

\noindent where $n_{OVIII, flat}$ and $n_{OVII, flat}$ are the best-fit models presented in Table~\ref{table.mcmc_results} for an optically thin plasma.  The $n_{OVIII, grad}$ and $n_{OVII, grad}$ terms cancel out because we assume the \ion{O}{8} and \ion{O}{7} observations come from the same plasma (i.e., the plasmas are cospatial).  Thus, we take the ratio on the left side of Equation (\ref{eq.t_grad}) and map it into a temperature gradient $T_{grad}(r)$.  Figure~\ref{figure.temp_grad} shows these temperature corrections create a relatively small change in temperature with $T$ decreasing from $2.4 - 1.5 \times 10^6$ K from 1 - 250 kpc (the $T \propto r^{-0.08}$ line is only included for illustrative purposes).  This new temperature gradient changes our best-fit density profiles to make the \ion{O}{7} and \ion{O}{8} fits consistent with each other based on Equation (\ref{eq.dI_conserved}).  The corresponding density profile is consistent with the initial \ion{O}{7} and \ion{O}{8} fit results within $\approx 10$ kpc and remains consistent with the \ion{O}{7} fit results beyond $\approx 10$ kpc (see shaded regions in Figure~\ref{figure.temp_grad}).  This implies the corrected density profile predicts similar masses, cooling times, cooling rates, etc. as our \ion{O}{7} flat temperature profile fit results ($5-9 \times 10^{10} M_{\odot}$ within 250 kpc).  We also note these corrected density and temperature profiles are significantly different from adiabatic profiles in hydrostatic equilibrium with the Milky Way's DM distribution used by \cite{maller_bullock04} and \cite{fang_etal13} to analyze the structure of the Milky Way's hot gas halo (maroon curves in Figure~\ref{figure.temp_grad}).  This inconsistency indicates there has been heating or cooling occurring in the halo gas.  

Given a new temperature and density profile for the halo gas, we calculate the entropy profile for the Milky Way's halo gas and compare it to the observed entropy profiles of galaxy groups and clusters.  Figure~\ref{figure.temp_grad} shows our model entropy profiles (defined as $S = kT/n_{e}^{2/3}$) scaled by the entropy at 0.3 $R_{vir}$.  The black and green shaded regions show the 1$\sigma$ boundary regions for the flat temperature and corrected temperature profiles respectively.  The orange shaded region represents the ``universal'' entropy profile for galaxy clusters determined from measuring the intracluster medium density and temperature profiles of approximately 15 galaxy clusters \citep{walker_etal12, okabe_etal14}.  This profile closely follows an $S \propto r^{1.1}$ slope (black dashed line) within 0.3 $R_{vir}$.  This slope is the characteristic entropy profile for gas accreting onto virialized objects in the absence of radiative cooling or additional heating due to feedback \citep{tozzi_norman01, voit_etal05}.  The yellow shaded region represents results from similar studies on samples of galaxy groups, with slopes ranging between $S \propto r^{0.5 - 0.7}$ \citep{finoguenov_etal07, panagoulia_etal14}.  Our corrected halo gas entropy profile is bounded by the scaled group and cluster profiles, and the profile uncertainties indicate that they are consistent with both profile shapes.  We also note the inconsistency between our profiles and the group/cluster profiles compared to the adiabatic profile from \cite{maller_bullock04} and \cite{fang_etal13}.  Our entropy profiles are consistent with the observed ``universal'' profiles around more massive virialized objects.  

The exercise discussed above is a simple estimate that yields a relatively minor temperature gradient compared to the observed halo gas temperature of $\approx 2 \times 10^6$ K (assuming the halo gas is isothermal).  We remind the reader that it is unclear if the \ion{O}{7} fit results are a valid description of the halo gas profile given the fits are still unacceptable at this point.  This caveat implies that we are missing some emission contribution in our model that has a much stronger effect in \ion{O}{7} line emission compared to \ion{O}{8} emission.  

We examine if SWCX emission contributes to the \ion{O}{7} - \ion{O}{8} $\chi_{red}^{2}$ discrepancy, although we conclude it is not the primary driver of the unacceptable \ion{O}{7} fits.  This is a possible source for this discrepancy since the typical SWCX \ion{O}{7} emission is larger than the typical \ion{O}{8} line emission \citep{koutroumpa_etal07}.  The relative strengths of the typical \ion{O}{7} versus \ion{O}{8} line emission due to SWCX reactions may cause many of the \ion{O}{7} observations to deviate from our smooth model predictions more than the \ion{O}{8} observations.  However, the sample we are analyzing was already subjected to a reasonable SWCX screening procedure outlined in \cite{hs12} (their flux-filtered sample).  This screening likely removed most of the geocoronal SWCX emission, which should correlate with the solar wind proton flux \citep{robertson_cravens03_a, robertson_cravens03_b, robertson_etal06}.  We also analyze a subset of observations near the ecliptic plane (see gray strip in Figure~\ref{figure.map_spacecut}) to probe heliospheric SWCX contamination.  This contamination is expected to be stronger near the ecliptic plane, although with longer temporal variation compared to geocoronal SWCX emission \citep{robertson_cravens03_a, koutroumpa_etal06}.  We find no noticeable excess in the emission line strengths, the strengths of the observed minus model residual emission, or the outlier strength (defined as the absolute value of the difference between the model and observed value divided by the measurement uncertainty) for sightlines near the ecliptic plane compared to sightlines in an equivalent gray strip rotated by 180$\arcdeg$ in Galactic longitude.  For example, the $\chi_{red}^{2}$ (dof) for the optically thin \ion{O}{7} observations is 5.35 (64) for the ecliptic plane strip and 5.39 (50) for the rotated ecliptic plane strip.  These lines of evidence imply that the \textit{global} properties of the emission line observations are not affected by SWCX emission, even for the \ion{O}{7} observations.  There is likely a different source for our \ion{O}{7} versus \ion{O}{8} fit quality discrepancy.  

The more likely explanation to our \ion{O}{7} - \ion{O}{8} fit discrepancy is variation in the physical properties of the X-ray-emitting gas creating these emission lines.  Specifically, variation in the density/temperature structure associated with the LB or inner regions of the hot gas halo can cause the observations to deviate from a smooth density profile.  Variation in the LB plasma is a likely scenario here because the ``local'' emission source (SWCX or LB) contributes more \ion{O}{7} than \ion{O}{8} emission (if any) compared to the ``non-local'' emission source.  This evidence comes from a range of shadowing experiments \citep[e.g.,][]{kuntz_snowden00, smith_etal07, koutroumpa_etal11} and from additional analyses on the \textit{ROSAT} R12 band maps \citep[e.g.,][]{galeazzi_etal14}.  These results indicate that regardless of the physical properties of the ``local'' X-ray-emitting gas, the emission is patchy across the sky.  On the other hand, the \ion{O}{7} absorption line analysis by \cite{miller_bregman13} showed they could only find an acceptable fit to the observations with the inclusion of an additional uncertainty of 7.5 m\AA\ to the measured equivalent widths ($\approx$30\% of the average equivalent width).  This variation is comparable to what we see with the \ion{O}{7} emission lines, where we must add an additional 2.1 L.U. uncertainty to the measurements ($\approx$40\% of the median \ion{O}{7} line intensity) to find an acceptable $\chi_{red}^{2}$.  Given these two types of analyses, we cannot definitively say which plasma causes the \ion{O}{7} absorption/emission line variation.  Understanding the source of the variation is beyond the scope of this work, but will involve detailed analysis on physical models of the local ISM \citep{deavillez_etal13}.

\section{Summary}
\label{section.summary}

We have presented an in-depth analysis of the Milky Way's SXRB using the largest sample of \ion{O}{7} and \ion{O}{8} emission lines to date.  Our sample is a subset of the work by \citep{hs10, hs12}, who presented \ion{O}{7} and \ion{O}{8} emission line measurements of the SXRB from \textit{XMM-Newton} archival data.  We applied additional observation screening methods to their sample to maximize our sensitivity to emission from the Milky Way's hot gas halo.  These screening methods left us with 649 out of 1003 observations from the \cite{hs12} flux-filtered sample covering all regions of the sky outside the Galactic center and Galactic plane.  The combination of the size and sky coverage of this sample allows us to constrain the physical properties of the Milky Way's hot gas halo much better than previous works.  

The advantage of this work over previous studies on the SXRB is that our model fitting procedure to the emission lines is identical to the work by \cite{miller_bregman13} on \ion{O}{7} absorption lines.  This is critical since the both types of observations are likely due to the same plasma sources in the Milky Way (although the different sources are expected to have different strengths in absorption and emission).  Thus, this work is a positive step toward unifying the absorption and emission line observations associated with the Milky Way's ISM/CGM.  

We find an acceptable fit to the \ion{O}{8} observations with a diffuse volume-filled hot gas halo model described as a modified $\beta$-model (a power law).  Our best-fit fitting results depend on whether we account for optical depth effects in the plasma, but we constrain $\beta$ to be between 0.50 and 0.54 without and with these corrections.  We also include a simple parameterization for LB emission in our model, but its contribution to the \ion{O}{8} emission lines is negligible ($\lesssim$0.02 L.U.) and it is unconstrained by the observations.

The \ion{O}{7} observations show considerably different behavior than the \ion{O}{8} observations, both in the quality of our best-fit models and in the best-fit parameters themselves.  The best-fit $\beta$ parameter is consistently smaller (shallower) when we analyze the \ion{O}{7} observations compared to the \ion{O}{8} observations.  This is possibly due to departures from an isothermal halo profile assumed in the analysis.  This interpretation is speculative however, since our best-fit parameters to the \ion{O}{7} observations yield unacceptable $\chi_{red}^{2}$ (dof) = 4.7 (645).  This implies there is significant sightline-to-sightline variation in the \ion{O}{7} observations that deviate the observations from our smooth hot halo + LB emission model.  

The implications of our model-fitting results are discussed in detail in Section ~\ref{section.discussion}.  We reiterate the most important results here.  

\begin{enumerate}

\item{The \ion{O}{8} model fitting results are consistent with previous work on \ion{O}{7} absorption lines utilizing the same model fitting procedure \citep{miller_bregman13}.  The fact these results are consistent with each other implies we are starting to develop a cohesive picture of the Milky Way's hot gas halo structure.  Specifically, this result is a positive step toward unifying the emission and absorption line observations that are due to the Milky Way's hot gas halo.  }

\item{The inferred mass from our \ion{O}{8} best-fit hot gas halo parameters ranges between $2.7 - 4.7 \times 10^{10} M_{\odot}$ within 250 kpc.  This mass is considerable when compared to the known baryon mass in the Milky Way ($6-7 \times 10^{10} M_{\odot}$), but is 6-10 times smaller than the missing baryon mass in the Milky Way.   }

\item{Several computed quantities from our best-fit model results suggest a gas metallicity of 0.3 $Z_{\odot}$.  The halo gas metallicity must be $\geq$0.3 $Z_{\odot}$ to not overproduce the residual pulsar DM toward the LMC due to a hot gas halo component.  We also are consistent with the previously observed 0.5-2.0 X-ray luminosity for the Milky Way \citep[$\sim 2 \times 10^{39}$ erg s$^{-1}$;][]{snowden_etal97} if we assume a gas metallicity of 0.3 $Z_{\odot}$.  An independent estimate of the halo gas metallicity using absorption line profile results from \cite{miller_bregman13} and these current emission line results also suggests a gas metallicity of $\approx$0.3 $Z_{\odot}$.  This metallicity is also consistent with simulations of galactic coronae \citep{toft_etal02, cen_ostriker06, cen12}.  }

\item{The discrepancy between our \ion{O}{8} and \ion{O}{7} fit results is likely due to variation in the emission properties of the LB rather than residual SWCX emission.  This patchiness in the emission has been analyzed in the ROSAT R12 band \citep[1/4 keV;][]{kuntz_snowden00, galeazzi_etal14}, which would have a stronger effect on the \ion{O}{7} emission lines compared to the \ion{O}{8} emission lines.  We attempt to quantify this patchiness in the emission lines by adding an uncertainty to the \ion{O}{7} observations.  We find we must add an uncertainty of 2.1 L.U. to the emission lines ($\approx$40\% the median line strength) to find an acceptable $\chi_{red}^{2}$.  }

\item{Optical depth corrections are likely necessary in our model emission line calculation since evidence suggests the hot halo plasma has optical depths of order unity \citep{williams_etal05, gupta_etal12, miller_bregman13}.  We attempt to quantify these effects and find optical depth corrections to the emission line calculations increase (steepen) our fitted $\beta$ by about 10\%. }

\end{enumerate}

Future work will involve rectifying the final two topics discussed above.  We aim to reproduce the sightline-to-sightline variability in the \ion{O}{7} observations with a physical model plasma model including our smooth hot gas halo profile with a presumably more clumpy or variable LB model.  This will involve detailed observational and theoretical work on the local ISM/CGM.

We also intend to utilize more detailed radiative transfer codes to quantify the optical depth effects present in the plasma.  Our simple parameterization of the optical depth corrections indicates these effects may be minor, but they will provide even tighter constraints on the Milky Way's hot gas density profile.  This will lead to improved estimates on quantities such as the mass and metallicity of the halo gas.  

Even with the above limitations, we emphasize the significant improvement these results are compared to previous work on the Milky Way's hot gas halo.  Not only have we dramatically reduced the density and resultant mass uncertainties for the Milky Way's hot gas halo, but have done so while utilizing emission lines rather than absorption lines.  This supports the power of these new large samples of emission lines (several hundred sightlines) compared with the much smaller absorption line samples ($\approx$30 sightlines).  We intend to use these emission line observations and modeling techniques for future work on other Galactic-scale features observed in X-rays.


\acknowledgments

The authors thank Michael Anderson, David Henley, Edmund Hodges-Kluck, and Sebasti\'{a}n Nuza for helpful discussions and comments on this work.  We also thank the anonymous referee for insightful comments that improved our work.  This research is based on observations obtained with \textit{XMM-Newton}, an ESA science mission with instruments and contributions directly funded by ESA Member States and NASA.  The authors gratefully acknowledge financial support from NASA through ADAP grant NNX11AJ55G.  

\bibliographystyle{apj}

\begin{thebibliography}{}
\expandafter\ifx\csname natexlab\endcsname\relax\def\natexlab#1{#1}\fi

\bibitem[{{Anders} \& {Grevesse}(1989)}]{anders_grevesse89}
{Anders}, E., \& {Grevesse}, N. 1989, \gca, 53, 197

\bibitem[{{Anderson} \& {Bregman}(2010)}]{anderson_bregman10}
{Anderson}, M.~E., \& {Bregman}, J.~N. 2010, \apj, 714, 320

\bibitem[{{Anderson} \& {Bregman}(2011)}]{anderson_bregman11}
---. 2011, \apj, 737, 22

\bibitem[{{Asplund} {et~al.}(2005){Asplund}, {Grevesse}, \&
  {Sauval}}]{asplund_etal05}
{Asplund}, M., {Grevesse}, N., \& {Sauval}, A.~J. 2005, in Astronomical Society
  of the Pacific Conference Series, Vol. 336, Cosmic Abundances as Records of
  Stellar Evolution and Nucleosynthesis, ed. T.~G. {Barnes}, III \& F.~N.
  {Bash}, 25

\bibitem[{{Balucinska-Church} \& {McCammon}(1992)}]{bc_mccammon92}
{Balucinska-Church}, M., \& {McCammon}, D. 1992, \apj, 400, 699

\bibitem[{{Binney} \& {Tremaine}(2008)}]{binney_tremaine08}
{Binney}, J., \& {Tremaine}, S. 2008, {Galactic Dynamics: Second Edition}
  (Princeton University Press)

\bibitem[{{Bland-Hawthorn} \& {Cohen}(2003)}]{bh_cohen03}
{Bland-Hawthorn}, J., \& {Cohen}, M. 2003, \apj, 582, 246

\bibitem[{{Boylan-Kolchin} {et~al.}(2013){Boylan-Kolchin}, {Bullock}, {Sohn},
  {Besla}, \& {van der Marel}}]{boylan_kolchin_etal13}
{Boylan-Kolchin}, M., {Bullock}, J.~S., {Sohn}, S.~T., {Besla}, G., \& {van der
  Marel}, R.~P. 2013, \apj, 768, 140

\bibitem[{{Bregman} {et~al.}(2015){Bregman}, {Anderson}, \&
  {Miller}}]{bregman_etal15}
{Bregman}, J.~N., {Anderson}, M.~E., \& {Miller}, M.~J. 2015, in preparation

\bibitem[{{Bregman} \& {Lloyd-Davies}(2007)}]{bregman_ld07}
{Bregman}, J.~N., \& {Lloyd-Davies}, E.~J. 2007, \apj, 669, 990

\bibitem[{{Brinkmann} {et~al.}(1997){Brinkmann}, {Yuan}, \&
  {Siebert}}]{brinkman_etal97}
{Brinkmann}, W., {Yuan}, W., \& {Siebert}, J. 1997, \aap, 319, 413

\bibitem[{{Carter} \& {Sembay}(2008)}]{carter_sembay08}
{Carter}, J.~A., \& {Sembay}, S. 2008, \aap, 489, 837

\bibitem[{{Carter} {et~al.}(2011){Carter}, {Sembay}, \& {Read}}]{carter_etal11}
{Carter}, J.~A., {Sembay}, S., \& {Read}, A.~M. 2011, \aap, 527, A115

\bibitem[{{Cen}(2012)}]{cen12}
{Cen}, R. 2012, \apj, 753, 17

\bibitem[{{Cen} \& {Ostriker}(2006)}]{cen_ostriker06}
{Cen}, R., \& {Ostriker}, J.~P. 2006, \apj, 650, 560

\bibitem[{{Chen} {et~al.}(1997){Chen}, {Fabian}, \& {Gendreau}}]{chen_etal97}
{Chen}, L.-W., {Fabian}, A.~C., \& {Gendreau}, K.~C. 1997, \mnras, 285, 449

\bibitem[{{Cravens} {et~al.}(2001){Cravens}, {Robertson}, \&
  {Snowden}}]{cravens_etal01}
{Cravens}, T.~E., {Robertson}, I.~P., \& {Snowden}, S.~L. 2001, \jgr, 106,
  24883

\bibitem[{{Dai} {et~al.}(2012){Dai}, {Anderson}, {Bregman}, \&
  {Miller}}]{dai_etal12}
{Dai}, X., {Anderson}, M.~E., {Bregman}, J.~N., \& {Miller}, J.~M. 2012, \apj,
  755, 107

\bibitem[{{de Avillez} {et~al.}(2013){de Avillez}, {Breitschwerdt}, {Asgekar},
  \& {Spitoni}}]{deavillez_etal13}
{de Avillez}, M.~A., {Breitschwerdt}, D., {Asgekar}, A., \& {Spitoni}, E. 2013,
  ArXiv e-prints, arXiv:1301.2890

\bibitem[{{de Avillez} \& {Mac Low}(2001)}]{deavillez_maclow01}
{de Avillez}, M.~A., \& {Mac Low}, M.-M. 2001, \apjl, 551, L57

\bibitem[{{Dunkley} {et~al.}(2009){Dunkley}, {Komatsu}, {Nolta}, {Spergel},
  {Larson}, {Hinshaw}, {Page}, {Bennett}, {Gold}, {Jarosik}, {Weiland},
  {Halpern}, {Hill}, {Kogut}, {Limon}, {Meyer}, {Tucker}, {Wollack}, \&
  {Wright}}]{dunkley_etal09}
{Dunkley}, J., {Komatsu}, E., {Nolta}, M.~R., {et~al.} 2009, \apjs, 180, 306

\bibitem[{{Ezoe} {et~al.}(2010){Ezoe}, {Ebisawa}, {Yamasaki}, {Mitsuda},
  {Yoshitake}, {Terada}, {Miyoshi}, \& {Fujimoto}}]{ezoe_etal10}
{Ezoe}, Y., {Ebisawa}, K., {Yamasaki}, N.~Y., {et~al.} 2010, \pasj, 62, 981

\bibitem[{{Fang} {et~al.}(2013){Fang}, {Bullock}, \&
  {Boylan-Kolchin}}]{fang_etal13}
{Fang}, T., {Bullock}, J., \& {Boylan-Kolchin}, M. 2013, \apj, 762, 20

\bibitem[{{Fang} \& {Jiang}(2014)}]{fang_jiang14}
{Fang}, T., \& {Jiang}, X. 2014, \apjl, 785, L24

\bibitem[{{Fang} {et~al.}(2006){Fang}, {Mckee}, {Canizares}, \&
  {Wolfire}}]{fang_etal06}
{Fang}, T., {Mckee}, C.~F., {Canizares}, C.~R., \& {Wolfire}, M. 2006, \apj,
  644, 174

\bibitem[{{Feldmann} {et~al.}(2013){Feldmann}, {Hooper}, \&
  {Gnedin}}]{feldmann_etal13}
{Feldmann}, R., {Hooper}, D., \& {Gnedin}, N.~Y. 2013, \apj, 763, 21

\bibitem[{{Finoguenov} {et~al.}(2007){Finoguenov}, {Ponman}, {Osmond}, \&
  {Zimer}}]{finoguenov_etal07}
{Finoguenov}, A., {Ponman}, T.~J., {Osmond}, J.~P.~F., \& {Zimer}, M. 2007,
  \mnras, 374, 737

\bibitem[{{Foreman-Mackey} {et~al.}(2013){Foreman-Mackey}, {Hogg}, {Lang}, \&
  {Goodman}}]{foreman_mackey_etal13}
{Foreman-Mackey}, D., {Hogg}, D.~W., {Lang}, D., \& {Goodman}, J. 2013, \pasp,
  125, 306

\bibitem[{{Forman} {et~al.}(1985){Forman}, {Jones}, \&
  {Tucker}}]{forman_etal85}
{Forman}, W., {Jones}, C., \& {Tucker}, W. 1985, \apj, 293, 102

\bibitem[{{Foster} {et~al.}(2012){Foster}, {Ji}, {Smith}, \&
  {Brickhouse}}]{foster_etal12}
{Foster}, A.~R., {Ji}, L., {Smith}, R.~K., \& {Brickhouse}, N.~S. 2012, \apj,
  756, 128

\bibitem[{{Fraternali} \& {Binney}(2008)}]{fraternali_binney08}
{Fraternali}, F., \& {Binney}, J.~J. 2008, \mnras, 386, 935

\bibitem[{{Fukugita} \& {Peebles}(2006)}]{fukugita_peebles06}
{Fukugita}, M., \& {Peebles}, P.~J.~E. 2006, \apj, 639, 590

\bibitem[{{Galeazzi} {et~al.}(2007){Galeazzi}, {Gupta}, {Covey}, \&
  {Ursino}}]{galeazzi_etal07}
{Galeazzi}, M., {Gupta}, A., {Covey}, K., \& {Ursino}, E. 2007, \apj, 658, 1081

\bibitem[{{Galeazzi} {et~al.}(2014){Galeazzi}, {Chiao}, {Collier}, {Cravens},
  {Koutroumpa}, {Kuntz}, {Lallement}, {Lepri}, {McCammon}, {Morgan}, {Porter},
  {Robertson}, {Snowden}, {Thomas}, {Uprety}, {Ursino}, \&
  {Walsh}}]{galeazzi_etal14}
{Galeazzi}, M., {Chiao}, M., {Collier}, M.~R., {et~al.} 2014, ArXiv e-prints,
  arXiv:1407.7539

\bibitem[{{Goodman} \& {Weare}(2010)}]{goodman_weare10}
{Goodman}, J., \& {Weare}, J. 2010, Comm. App. Math. Comp. Sci., 5, 65

\bibitem[{{Gupta} {et~al.}(2012){Gupta}, {Mathur}, {Krongold}, {Nicastro}, \&
  {Galeazzi}}]{gupta_etal12}
{Gupta}, A., {Mathur}, S., {Krongold}, Y., {Nicastro}, F., \& {Galeazzi}, M.
  2012, \apjl, 756, L8

\bibitem[{{Hagihara} {et~al.}(2010){Hagihara}, {Yao}, {Yamasaki}, {Mitsuda},
  {Wang}, {Takei}, {Yoshino}, \& {McCammon}}]{hagihara_etal10}
{Hagihara}, T., {Yao}, Y., {Yamasaki}, N.~Y., {et~al.} 2010, \pasj, 62, 723

\bibitem[{{Henley} \& {Shelton}(2010)}]{hs10}
{Henley}, D.~B., \& {Shelton}, R.~L. 2010, \apjs, 187, 388

\bibitem[{{Henley} \& {Shelton}(2012)}]{hs12}
---. 2012, \apjs, 202, 14

\bibitem[{{Henley} \& {Shelton}(2013)}]{hs13}
---. 2013, \apj, 773, 92

\bibitem[{{Henley} {et~al.}(2010){Henley}, {Shelton}, {Kwak}, {Joung}, \& {Mac
  Low}}]{henley_etal10}
{Henley}, D.~B., {Shelton}, R.~L., {Kwak}, K., {Joung}, M.~R., \& {Mac Low},
  M.-M. 2010, \apj, 723, 935

\bibitem[{{Hill} {et~al.}(2012){Hill}, {Joung}, {Mac Low}, {Benjamin},
  {Haffner}, {Klingenberg}, \& {Waagan}}]{hill_etal12}
{Hill}, A.~S., {Joung}, M.~R., {Mac Low}, M.-M., {et~al.} 2012, \apj, 750, 104

\bibitem[{{Holweger}(2001)}]{holweger_01}
{Holweger}, H. 2001, in American Institute of Physics Conference Series, Vol.
  598, Joint SOHO/ACE workshop ''Solar and Galactic Composition'', ed. R.~F.
  {Wimmer-Schweingruber}, 23--30

\bibitem[{{Joung} \& {Mac Low}(2006)}]{joung_maclow06}
{Joung}, M.~K.~R., \& {Mac Low}, M.-M. 2006, \apj, 653, 1266

\bibitem[{{Kalberla} {et~al.}(2005){Kalberla}, {Burton}, {Hartmann}, {Arnal},
  {Bajaja}, {Morras}, \& {P{\"o}ppel}}]{kalberla_etal05}
{Kalberla}, P.~M.~W., {Burton}, W.~B., {Hartmann}, D., {et~al.} 2005, \aap,
  440, 775

\bibitem[{{Kataoka} {et~al.}(2013){Kataoka}, {Tahara}, {Totani}, {Sofue},
  {Stawarz}, {Takahashi}, {Takeuchi}, {Tsunemi}, {Kimura}, {Takei}, {Cheung},
  {Inoue}, \& {Nakamori}}]{kataoka_etal13}
{Kataoka}, J., {Tahara}, M., {Totani}, T., {et~al.} 2013, \apj, 779, 57

\bibitem[{{Keshet} {et~al.}(2004){Keshet}, {Waxman}, \& {Loeb}}]{keshet_etal04}
{Keshet}, U., {Waxman}, E., \& {Loeb}, A. 2004, JCAP, 4, 6

\bibitem[{{Klypin} {et~al.}(2002){Klypin}, {Zhao}, \&
  {Somerville}}]{klypin_etal02}
{Klypin}, A., {Zhao}, H., \& {Somerville}, R.~S. 2002, \apj, 573, 597

\bibitem[{{Koutroumpa} {et~al.}(2007){Koutroumpa}, {Acero}, {Lallement},
  {Ballet}, \& {Kharchenko}}]{koutroumpa_etal07}
{Koutroumpa}, D., {Acero}, F., {Lallement}, R., {Ballet}, J., \& {Kharchenko},
  V. 2007, \aap, 475, 901

\bibitem[{{Koutroumpa} {et~al.}(2006){Koutroumpa}, {Lallement}, {Kharchenko},
  {Dalgarno}, {Pepino}, {Izmodenov}, \& {Qu{\'e}merais}}]{koutroumpa_etal06}
{Koutroumpa}, D., {Lallement}, R., {Kharchenko}, V., {et~al.} 2006, \aap, 460,
  289

\bibitem[{{Koutroumpa} {et~al.}(2011){Koutroumpa}, {Smith}, {Edgar}, {Kuntz},
  {Plucinsky}, \& {Snowden}}]{koutroumpa_etal11}
{Koutroumpa}, D., {Smith}, R.~K., {Edgar}, R.~J., {et~al.} 2011, \apj, 726, 91

\bibitem[{{Kuntz} \& {Snowden}(2000)}]{kuntz_snowden00}
{Kuntz}, K.~D., \& {Snowden}, S.~L. 2000, \apj, 543, 195

\bibitem[{{Kuntz} \& {Snowden}(2008)}]{kuntz_snowden08}
---. 2008, \aap, 478, 575

\bibitem[{{Lallement} {et~al.}(2003){Lallement}, {Welsh}, {Vergely}, {Crifo},
  \& {Sfeir}}]{lallement_etal03}
{Lallement}, R., {Welsh}, B.~Y., {Vergely}, J.~L., {Crifo}, F., \& {Sfeir}, D.
  2003, \aap, 411, 447

\bibitem[{{Lei} {et~al.}(2009){Lei}, {Shelton}, \& {Henley}}]{lei_etal09}
{Lei}, S., {Shelton}, R.~L., \& {Henley}, D.~B. 2009, \apj, 699, 1891

\bibitem[{{Leitner} \& {Kravtsov}(2011)}]{leitner_kravtsov11}
{Leitner}, S.~N., \& {Kravtsov}, A.~V. 2011, \apj, 734, 48

\bibitem[{{Loeb} {et~al.}(2005){Loeb}, {Reid}, {Brunthaler}, \&
  {Falcke}}]{loeb_etal05}
{Loeb}, A., {Reid}, M.~J., {Brunthaler}, A., \& {Falcke}, H. 2005, \apj, 633,
  894

\bibitem[{{Mac Low} {et~al.}(1989){Mac Low}, {McCray}, \&
  {Norman}}]{maclow_etal89}
{Mac Low}, M.-M., {McCray}, R., \& {Norman}, M.~L. 1989, \apj, 337, 141

\bibitem[{{Maller} \& {Bullock}(2004)}]{maller_bullock04}
{Maller}, A.~H., \& {Bullock}, J.~S. 2004, \mnras, 355, 694

\bibitem[{{Marinacci} {et~al.}(2014){Marinacci}, {Pakmor}, {Springel}, \&
  {Simpson}}]{marinacci_etal14}
{Marinacci}, F., {Pakmor}, R., {Springel}, V., \& {Simpson}, C.~M. 2014,
  \mnras, 442, 3745

\bibitem[{{McCammon} {et~al.}(2002){McCammon}, {Almy}, {Apodaca}, {Bergmann
  Tiest}, {Cui}, {Deiker}, {Galeazzi}, {Juda}, {Lesser}, {Mihara},
  {Morgenthaler}, {Sanders}, {Zhang}, {Figueroa-Feliciano}, {Kelley},
  {Moseley}, {Mushotzky}, {Porter}, {Stahle}, \&
  {Szymkowiak}}]{mccammon_etal02}
{McCammon}, D., {Almy}, R., {Apodaca}, E., {et~al.} 2002, \apj, 576, 188

\bibitem[{{Mewe} {et~al.}(1995){Mewe}, {Kaastra}, \& {Liedahl}}]{mewe_etal95}
{Mewe}, R., {Kaastra}, J.~S., \& {Liedahl}, D.~A. 1995, Legacy, 6, 16

\bibitem[{{Miller} {et~al.}(2008){Miller}, {Tsunemi Hiroshi}, {Bautz},
  {McCammon}, {Fujimoto}, {Hughes}, {Katsuda}, {Kokubun}, {Mitsuda}, {Porter},
  {Takei}, {Tsuboi}, \& {Yamasaki}}]{miller_etal08}
{Miller}, E.~D., {Tsunemi Hiroshi}, {Bautz}, M.~W., {et~al.} 2008, \pasj, 60,
  95

\bibitem[{{Miller} \& {Bregman}(2013)}]{miller_bregman13}
{Miller}, M.~J., \& {Bregman}, J.~N. 2013, \apj, 770, 118

\bibitem[{{Moretti} {et~al.}(2003){Moretti}, {Campana}, {Lazzati}, \&
  {Tagliaferri}}]{moretti_etal03}
{Moretti}, A., {Campana}, S., {Lazzati}, D., \& {Tagliaferri}, G. 2003, \apj,
  588, 696

\bibitem[{{Nicastro} {et~al.}(2002){Nicastro}, {Zezas}, {Drake}, {Elvis},
  {Fiore}, {Fruscione}, {Marengo}, {Mathur}, \& {Bianchi}}]{nicastro_etal02}
{Nicastro}, F., {Zezas}, A., {Drake}, J., {et~al.} 2002, \apj, 573, 157

\bibitem[{{Norman} \& {Ikeuchi}(1989)}]{norman_ikeuchi89}
{Norman}, C.~A., \& {Ikeuchi}, S. 1989, \apj, 345, 372

\bibitem[{{Nuza} {et~al.}(2014){Nuza}, {Parisi}, {Scannapieco}, {Richter},
  {Gottl{\"o}ber}, \& {Steinmetz}}]{nuza_etal14}
{Nuza}, S.~E., {Parisi}, F., {Scannapieco}, C., {et~al.} 2014, \mnras, 441,
  2593

\bibitem[{{Okabe} {et~al.}(2014){Okabe}, {Umetsu}, {Tamura}, {Fujita},
  {Takizawa}, {Zhang}, {Matsushita}, {Hamana}, {Fukazawa}, {Futamase},
  {Kawaharada}, {Miyazaki}, {Mochizuki}, {Nakazawa}, {Ohashi}, {Ota}, {Sasaki},
  {Sato}, \& {Tam}}]{okabe_etal14}
{Okabe}, N., {Umetsu}, K., {Tamura}, T., {et~al.} 2014, \pasj,
  doi:10.1093/pasj/psu075

\bibitem[{{O'Sullivan} {et~al.}(2003){O'Sullivan}, {Ponman}, \&
  {Collins}}]{osullivan_etal03}
{O'Sullivan}, E., {Ponman}, T.~J., \& {Collins}, R.~S. 2003, \mnras, 340, 1375

\bibitem[{{Paerels} \& {Kahn}(2003)}]{paerels_kahn03}
{Paerels}, F.~B.~S., \& {Kahn}, S.~M. 2003, \araa, 41, 291

\bibitem[{{Panagoulia} {et~al.}(2014){Panagoulia}, {Fabian}, \&
  {Sanders}}]{panagoulia_etal14}
{Panagoulia}, E.~K., {Fabian}, A.~C., \& {Sanders}, J.~S. 2014, \mnras, 438,
  2341

\bibitem[{{Paturel} {et~al.}(2003){Paturel}, {Petit}, {Prugniel}, {Theureau},
  {Rousseau}, {Brouty}, {Dubois}, \& {Cambr{\'e}sy}}]{paturel_etal03}
{Paturel}, G., {Petit}, C., {Prugniel}, P., {et~al.} 2003, \aap, 412, 45

\bibitem[{{Piffaretti} {et~al.}(2011){Piffaretti}, {Arnaud}, {Pratt},
  {Pointecouteau}, \& {Melin}}]{piffaretti_etal11}
{Piffaretti}, R., {Arnaud}, M., {Pratt}, G.~W., {Pointecouteau}, E., \&
  {Melin}, J.-B. 2011, \aap, 534, A109

\bibitem[{{Rasmussen} {et~al.}(2003){Rasmussen}, {Kahn}, \&
  {Paerels}}]{rasmussen_etal03}
{Rasmussen}, A., {Kahn}, S.~M., \& {Paerels}, F. 2003, in Astrophysics and
  Space Science Library, Vol. 281, The IGM/Galaxy Connection. The Distribution
  of Baryons at z=0, ed. J.~L. {Rosenberg} \& M.~E. {Putman}, 109

\bibitem[{{Raymond} \& {Smith}(1977)}]{raymond_smith77}
{Raymond}, J.~C., \& {Smith}, B.~W. 1977, \apjs, 35, 419

\bibitem[{{Robertson} {et~al.}(2006){Robertson}, {Collier}, {Cravens}, \&
  {Fok}}]{robertson_etal06}
{Robertson}, I.~P., {Collier}, M.~R., {Cravens}, T.~E., \& {Fok}, M.-C. 2006,
  Journal of Geophysical Research (Space Physics), 111, 12105

\bibitem[{{Robertson} \& {Cravens}(2003{\natexlab{a}})}]{robertson_cravens03_a}
{Robertson}, I.~P., \& {Cravens}, T.~E. 2003{\natexlab{a}}, Journal of
  Geophysical Research (Space Physics), 108, 8031

\bibitem[{{Robertson} \& {Cravens}(2003{\natexlab{b}})}]{robertson_cravens03_b}
---. 2003{\natexlab{b}}, \grl, 30, 1439

\bibitem[{{Robitaille} \& {Whitney}(2010)}]{robitaille_whitney10}
{Robitaille}, T.~P., \& {Whitney}, B.~A. 2010, \apjl, 710, L11

\bibitem[{{Shattow} \& {Loeb}(2009)}]{shattow_loeb09}
{Shattow}, G., \& {Loeb}, A. 2009, \mnras, 392, L21

\bibitem[{{Smith} {et~al.}(2001){Smith}, {Brickhouse}, {Liedahl}, \&
  {Raymond}}]{smith_etal01}
{Smith}, R.~K., {Brickhouse}, N.~S., {Liedahl}, D.~A., \& {Raymond}, J.~C.
  2001, \apjl, 556, L91

\bibitem[{{Smith} {et~al.}(2014){Smith}, {Foster}, {Edgar}, \&
  {Brickhouse}}]{smith_etal14}
{Smith}, R.~K., {Foster}, A.~R., {Edgar}, R.~J., \& {Brickhouse}, N.~S. 2014,
  \apj, 787, 77

\bibitem[{{Smith} {et~al.}(2007){Smith}, {Bautz}, {Edgar}, {Fujimoto},
  {Hamaguchi}, {Hughes}, {Ishida}, {Kelley}, {Kilbourne}, {Kuntz}, {McCammon},
  {Miller}, {Mitsuda}, {Mukai}, {Plucinsky}, {Porter}, {Snowden}, {Takei},
  {Terada}, {Tsuboi}, \& {Yamasaki}}]{smith_etal07}
{Smith}, R.~K., {Bautz}, M.~W., {Edgar}, R.~J., {et~al.} 2007, \pasj, 59, 141

\bibitem[{{Snowden} {et~al.}(2004){Snowden}, {Collier}, \&
  {Kuntz}}]{snowden_etal04}
{Snowden}, S.~L., {Collier}, M.~R., \& {Kuntz}, K.~D. 2004, \apj, 610, 1182

\bibitem[{{Snowden} {et~al.}(1990){Snowden}, {Cox}, {McCammon}, \&
  {Sanders}}]{snowden_etal90}
{Snowden}, S.~L., {Cox}, D.~P., {McCammon}, D., \& {Sanders}, W.~T. 1990, \apj,
  354, 211

\bibitem[{{Snowden} \& {Kuntz}(2011)}]{snowden_kuntz11}
{Snowden}, S.~L., \& {Kuntz}, K.~D. 2011, Cookbook for Analysis Procedures for
  XMM-Newton EPIC MOS Observations of Extended Objects and the Diffuse
  Background, version 4.3

\bibitem[{{Snowden} {et~al.}(1993){Snowden}, {McCammon}, \&
  {Verter}}]{snowden_etal93}
{Snowden}, S.~L., {McCammon}, D., \& {Verter}, F. 1993, \apjl, 409, L21

\bibitem[{{Snowden} {et~al.}(1997){Snowden}, {Egger}, {Freyberg}, {McCammon},
  {Plucinsky}, {Sanders}, {Schmitt}, {Tr{\"u}mper}, \&
  {Voges}}]{snowden_etal97}
{Snowden}, S.~L., {Egger}, R., {Freyberg}, M.~J., {et~al.} 1997, \apj, 485, 125

\bibitem[{{Sommer-Larsen}(2006)}]{sommer_larsen06}
{Sommer-Larsen}, J. 2006, \apjl, 644, L1

\bibitem[{{Stewart} {et~al.}(2013){Stewart}, {Brooks}, {Bullock}, {Maller},
  {Diemand}, {Wadsley}, \& {Moustakas}}]{stewart_etal13}
{Stewart}, K.~R., {Brooks}, A.~M., {Bullock}, J.~S., {et~al.} 2013, \apj, 769,
  74

\bibitem[{{Su} {et~al.}(2010){Su}, {Slatyer}, \& {Finkbeiner}}]{su_etal10}
{Su}, M., {Slatyer}, T.~R., \& {Finkbeiner}, D.~P. 2010, \apj, 724, 1044

\bibitem[{{Sutherland} \& {Dopita}(1993)}]{sutherland_dopita93}
{Sutherland}, R.~S., \& {Dopita}, M.~A. 1993, \apjs, 88, 253

\bibitem[{{Toft} {et~al.}(2002){Toft}, {Rasmussen}, {Sommer-Larsen}, \&
  {Pedersen}}]{toft_etal02}
{Toft}, S., {Rasmussen}, J., {Sommer-Larsen}, J., \& {Pedersen}, K. 2002,
  \mnras, 335, 799

\bibitem[{{Tozzi} \& {Norman}(2001)}]{tozzi_norman01}
{Tozzi}, P., \& {Norman}, C. 2001, \apj, 546, 63

\bibitem[{{Voges} {et~al.}(1999){Voges}, {Aschenbach}, {Boller},
  {Br{\"a}uninger}, {Briel}, {Burkert}, {Dennerl}, {Englhauser}, {Gruber},
  {Haberl}, {Hartner}, {Hasinger}, {K{\"u}rster}, {Pfeffermann}, {Pietsch},
  {Predehl}, {Rosso}, {Schmitt}, {Tr{\"u}mper}, \& {Zimmermann}}]{voges_etal99}
{Voges}, W., {Aschenbach}, B., {Boller}, T., {et~al.} 1999, \aap, 349, 389

\bibitem[{{Voit} {et~al.}(2005){Voit}, {Kay}, \& {Bryan}}]{voit_etal05}
{Voit}, G.~M., {Kay}, S.~T., \& {Bryan}, G.~L. 2005, \mnras, 364, 909

\bibitem[{{Walker} {et~al.}(2012){Walker}, {Fabian}, {Sanders}, \&
  {George}}]{walker_etal12}
{Walker}, S.~A., {Fabian}, A.~C., {Sanders}, J.~S., \& {George}, M.~R. 2012,
  \mnras, 427, L45

\bibitem[{{Wang}(1998)}]{wang98}
{Wang}, Q.~D. 1998, in Lecture Notes in Physics, Berlin Springer Verlag, Vol.
  506, IAU Colloq. 166: The Local Bubble and Beyond, ed. D.~{Breitschwerdt},
  M.~J. {Freyberg}, \& J.~{Truemper}, 503--512

\bibitem[{{Wang} {et~al.}(2005){Wang}, {Yao}, {Tripp}, {Fang}, {Cui},
  {Nicastro}, {Mathur}, {Williams}, {Song}, \& {Croft}}]{wang_etal05}
{Wang}, Q.~D., {Yao}, Y., {Tripp}, T.~M., {et~al.} 2005, \apj, 635, 386

\bibitem[{{Wargelin} {et~al.}(2004){Wargelin}, {Markevitch}, {Juda},
  {Kharchenko}, {Edgar}, \& {Dalgarno}}]{wargelin_etal04}
{Wargelin}, B.~J., {Markevitch}, M., {Juda}, M., {et~al.} 2004, \apj, 607, 596

\bibitem[{{Watson} {et~al.}(2009){Watson}, {Schr{\"o}der}, {Fyfe}, {Page},
  {Lamer}, {Mateos}, {Pye}, {Sakano}, {Rosen}, {Ballet}, {Barcons}, {Barret},
  {Boller}, {Brunner}, {Brusa}, {Caccianiga}, {Carrera}, {Ceballos}, {Della
  Ceca}, {Denby}, {Denkinson}, {Dupuy}, {Farrell}, {Fraschetti}, {Freyberg},
  {Guillout}, {Hambaryan}, {Maccacaro}, {Mathiesen}, {McMahon}, {Michel},
  {Motch}, {Osborne}, {Page}, {Pakull}, {Pietsch}, {Saxton}, {Schwope},
  {Severgnini}, {Simpson}, {Sironi}, {Stewart}, {Stewart}, {Stobbart}, {Tedds},
  {Warwick}, {Webb}, {West}, {Worrall}, \& {Yuan}}]{watson_etal09}
{Watson}, M.~G., {Schr{\"o}der}, A.~C., {Fyfe}, D., {et~al.} 2009, \aap, 493,
  339

\bibitem[{{Welsh} \& {Shelton}(2009)}]{welsh_shelton09}
{Welsh}, B.~Y., \& {Shelton}, R.~L. 2009, \apss, 323, 1

\bibitem[{{White} \& {Frenk}(1991)}]{white_frenk91}
{White}, S.~D.~M., \& {Frenk}, C.~S. 1991, \apj, 379, 52

\bibitem[{{Williams} {et~al.}(2005){Williams}, {Mathur}, {Nicastro}, {Elvis},
  {Drake}, {Fang}, {Fiore}, {Krongold}, {Wang}, \& {Yao}}]{williams_etal05}
{Williams}, R.~J., {Mathur}, S., {Nicastro}, F., {et~al.} 2005, \apj, 631, 856

\bibitem[{{Yan} {et~al.}(1998){Yan}, {Sadeghpour}, \& {Dalgarno}}]{yan_etal98}
{Yan}, M., {Sadeghpour}, H.~R., \& {Dalgarno}, A. 1998, \apj, 496, 1044

\bibitem[{{Yao} {et~al.}(2012){Yao}, {Shull}, {Wang}, \& {Cash}}]{yao_etal12}
{Yao}, Y., {Shull}, J.~M., {Wang}, Q.~D., \& {Cash}, W. 2012, \apj, 746, 166

\bibitem[{{Yao} \& {Wang}(2005)}]{yao_wang05}
{Yao}, Y., \& {Wang}, Q.~D. 2005, \apj, 624, 751

\bibitem[{{Yao} \& {Wang}(2007)}]{yao_wang07}
---. 2007, \apj, 658, 1088

\bibitem[{{Yao} {et~al.}(2009){Yao}, {Wang}, {Hagihara}, {Mitsuda}, {McCammon},
  \& {Yamasaki}}]{yao_etal09_b}
{Yao}, Y., {Wang}, Q.~D., {Hagihara}, T., {et~al.} 2009, \apj, 690, 143

\bibitem[{{Yoshino} {et~al.}(2009){Yoshino}, {Mitsuda}, {Yamasaki}, {Takei},
  {Hagihara}, {Masui}, {Bauer}, {McCammon}, {Fujimoto}, {Wang}, \&
  {Yao}}]{yoshino_etal09}
{Yoshino}, T., {Mitsuda}, K., {Yamasaki}, N.~Y., {et~al.} 2009, \pasj, 61, 805

\bibitem[{{Yuan} {et~al.}(1998){Yuan}, {Brinkmann}, {Siebert}, \&
  {Voges}}]{yuan_etal98}
{Yuan}, W., {Brinkmann}, W., {Siebert}, J., \& {Voges}, W. 1998, \aap, 330, 108

\end{thebibliography}


\clearpage


\clearpage

\begin{deluxetable}{l c c}

\tablewidth{0pt}
\tablecolumns{3}
\tablecaption{Automated Screening Criteria}
\tablehead{
  \colhead{Catalog} &
  \colhead{Types of Sources} &
  \colhead{ Screening Criteria \tablenotemark{a}}  \\
  \colhead{} &
  \colhead{} &
  \colhead{} 
  } 
  
\startdata
\label{table.screening_catalogs}

\textit{ROSAT}-BSC \tablenotemark{b} & Any bright X-ray source & \textgreater 1 counts s$^{-1}$ \\
\textit{ROSAT}-RLQ \tablenotemark{c} & Radio loud quasars      & $F_{.1-2.4}$ \tablenotemark{d} \textgreater 10$^{-11}$ erg cm$^{-2}$ s$^{-1}$ \\
\textit{ROSAT}-RQQ \tablenotemark{e} & Radio quiet quasars     & $F_{0.1-2.4}$ \textgreater 10$^{-11}$ erg cm$^{-2}$ s$^{-1}$ \\
PGC 2003           \tablenotemark{f} & Galaxies                & Apparent diameter \textgreater 10$\arcmin$ \\
MCXC               \tablenotemark{g} & Galaxy clusters         & z \textless .1
\enddata

\tablenotetext{a}{Objects satisfying these criteria in their respective catalogs compose our potential contaminant source list.  All observations from the \cite{hs12} Flux-filtered sample within 0.5$\arcdeg$ of these objects are removed in our model fitting procedure. }
\tablenotetext{b}{ROSAT All-Sky Survey Bright Source Catalog           \citep[\url{http://www.xray.mpe.mpg.de/rosat/survey/rass-bsc/}; ]        []{voges_etal99}.  }
\tablenotetext{c}{ROSAT Radio Loud Quasar Catalog                      \citep[\url{http://heasarc.gsfc.nasa.gov/W3Browse/rosat/rosatrlq.html}; ][]{brinkman_etal97}.  }
\tablenotetext{d}{0.1 - 2.4 keV flux.  }
\tablenotetext{e}{ROSAT Radio Quiet Quasar Catalog                     \citep[\url{http://heasarc.gsfc.nasa.gov/W3Browse/rosat/rosatrqq.html}; ][]{yuan_etal98}.  }
\tablenotetext{f}{Principal Galaxy Catalog                             \citep[\url{http://leda.univ-lyon1.fr/}; ]                               []{paturel_etal03}.  }
\tablenotetext{g}{Meta-Catalog of X-ray Detected Clusters of Galaxies  \citep[\url{http://heasarc.gsfc.nasa.gov/W3Browse/rosat/mcxc.html}; ]    []{piffaretti_etal11}.  }
\end{deluxetable}


\clearpage
\begin{turnpage}

\begin{deluxetable}{l l c c c c c c c}

\tablewidth{0pt}
\tablecolumns{9}
\tablecaption{Model Fitting Results}
\tablehead{
  \colhead{Lines Fitted}            &
  \colhead{Plasma Type \tablenotemark{a}}             &
  \colhead{$n_or_c^{3\beta}$}       &
  \colhead{$\beta$}                 &
  \colhead{$n_{LB}$}                &
  \colhead{$\sigma_{add}$ \tablenotemark{b}}          &
  \colhead{$\chi^{2}_{red}$ (dof)}  &
  \colhead{$M$(\textless 50 kpc)  \tablenotemark{c}}   &
  \colhead{$M$(\textless 250 kpc) \tablenotemark{c}}  \\
  \colhead{}                                     &
  \colhead{}                                     &
  \colhead{($10^{-2}$ cm$^{-3}$ kpc$^{3\beta}$)}   &
  \colhead{}                                     &
  \colhead{($10^{-3}$ cm$^{-3}$)}                  &
  \colhead{L.U.}                                 &
  \colhead{}                                     &
  \colhead{($10^9 M_{\odot}$)}                     &
  \colhead{($10^{10} M_{\odot}$)} 
  }  
\startdata

\label{table.mcmc_results}

\ion{O}{8}  & Optically thin                          & $1.35 \pm 0.24$ & $0.50 \pm 0.03$ & $0.77 \pm 4.10$ & None & 1.08 (644) & $3.8^{+0.3}_{-0.3}$  & $4.3^{+0.9}_{-0.8}$ \\
\ion{O}{8}  & $\tau$ corrections                      & $1.50 \pm 0.24$ & $0.54 \pm 0.03$ & $1.13 \pm 4.69$ & None & 1.08 (644) & $2.9^{+0.3}_{-0.4}$  & $2.7^{+0.7}_{-0.6}$ \\
\hline
\ion{O}{7}  & Optically thin                          & $0.89 \pm 0.06$ & $0.43 \pm 0.01$ & $3.86 \pm 0.26$ & None & 4.69 (645) & $5.1^{+0.2}_{-0.2}$  & $7.9^{+0.8}_{-0.8}$ \\
\ion{O}{7}  & Optically thin with $\sigma_{add}$      & $0.76 \pm 0.11$ & $0.41 \pm 0.03$ & $3.83 \pm 0.63$ & 2.1  & 1.03 (645) & $5.3^{+0.5}_{-0.6}$  & $9.1^{+2.2}_{-1.9}$ \\
\ion{O}{7}  & $\tau$ corrections                      & $0.91 \pm 0.06$ & $0.47 \pm 0.01$ & $4.02 \pm 0.25$ & None & 4.67 (645) & $3.5^{+0.2}_{-0.2}$  & $4.5^{+0.5}_{-0.5}$ \\
\ion{O}{7}  & $\tau$ corrections with $\sigma_{add}$  & $0.79 \pm 0.10$ & $0.45 \pm 0.03$ & $3.98 \pm 0.62$ & 2.1  & 1.03 (645) & $3.7^{+0.4}_{-0.4}$  & $5.2^{+1.4}_{-1.2}$
\enddata

\tablenotetext{a}{We calculate model line intensities assuming the plasma is optically thin or with optical depth corrections.}
\tablenotetext{b}{Added uncertainty to the observations required to find an acceptable $\chi^{2}_{red}$.}
\tablenotetext{c}{Hot gas masses assuming a gas metallicity of 0.3 $Z_{\odot}$.}
\end{deluxetable}

\end{turnpage}
\global\pdfpageattr\expandafter{\the\pdfpageattr/Rotate 90}
\clearpage

\end{document}